\documentclass[journal,twocolumn]{IEEEtran}
\usepackage{epsfig,rotating,setspace,latexsym,amsmath,epsf,amssymb,amsfonts,bm,theorem,cite,caption,subcaption,enumerate,longtable,accents,bbm,booktabs,multirow }
\usepackage{algorithm,algorithmic,graphicx,epsf,authblk,epstopdf,url,color,multirow}
\usepackage{mathtools,comment,tabularx}
\usepackage{comment}
\usepackage[inline]{enumitem}
\DeclarePairedDelimiter{\ceil}{\lceil}{\rceil}
\usepackage{dsfont}

\newtheorem{theorem}{Theorem}
\newtheorem{proposition}{Proposition}

\newtheorem{corollary}{Corollary}
\newtheorem{definition}{Definition}
\newtheorem{remark}{Remark}
\newtheorem{lemma}{Lemma}

\newenvironment{Proof}[1]{\medskip\par\noindent{\bf Proof:\,}\,#1}{{\mbox{\,$\blacksquare$}\par}}

\newcolumntype{Y}{>{\centering\arraybackslash}X}
\newcommand{\figref}[1]{\figurename~\ref{#1}}

\allowdisplaybreaks

\begin{document}

\title{When Should Selfish Miners Double-Spend?}

\author{Mustafa Doger \qquad Sennur Ulukus\\
\normalsize Department of Electrical and Computer Engineering\\
\normalsize University of Maryland, College Park, MD 20742\\
\normalsize  \emph{doger@umd.edu} \qquad \emph{ulukus@umd.edu}}

\maketitle

\begin{abstract}
     Conventional double-spending attack models ignore the revenue losses stemming from the orphan blocks. On the other hand, selfish mining literature usually ignores the chance of the attacker to double-spend at no-cost in each attack cycle. In this paper, we give a rigorous stochastic analysis of an attack where the goal of the adversary is to double-spend while mining selfishly. To do so, we first combine stubborn and selfish mining attacks,  \textit{i.e.}, construct a strategy where the attacker acts stubborn until its private branch reaches a certain length and then switches to act selfish. We provide the optimal stubbornness for each parameter regime. Next, we provide the maximum stubbornness that is still more profitable than honest mining and argue a connection between the level of stubbornness and the $k$-confirmation rule. We show that, at each attack cycle, if the level of stubbornness is higher than $k$, the adversary gets a free shot at double-spending. At each cycle, for a given stubbornness level, we rigorously formulate how great the probability of double-spending is. We further modify the attack in the stubborn regime in order to conceal the attack and increase the double-spending probability.
\end{abstract}


\section{Introduction}
After Nakamoto's seminal paper\cite{btc-whitepaper} that laid the foundations of blockchains aiming for large-scale consensus, there have been numerous works studying the related security guarantees and flaws \cite{blockchain_security_survey}. The private attack, investigated initially in the Bitcoin whitepaper \cite{btc-whitepaper}, aims to replace a confirmed block\footnote{A rule-of-thumb in Nakamoto blockchains is to treat a block that is $k$-block deep in the longest chain as a point of consensus, \textit{i.e.}, consider it as confirmed (\textit{e.g.}, $k=6$ for Bitcoin).} (hence, the transactions within) with a conflicting one, allowing the attacker to double-spend its funds. Rosenfeld \cite{rosenfeld2014analysis} showed rigorously that the private attack has exponentially decaying probability of success with time under the longest chain protocol. More specifically, as time passes, more public blocks are mined on top of a block that is being attacked, and it becomes less likely for a competing private chain to overtake the longest chain and undo the block in question, hence, a block that is sufficiently deep in the longest chain is a point of consensus.\footnote{Thus, double-spending is a poor operating procedure for a fixed $k$ of the $k$-confirmation rule which is a heuristic approach to reach consensus.}

The celebrated selfish mining attack of Eyal and Sirer \cite{selfish-mining} exposed another vulnerability of the longest chain protocol by showing that it is not incentive-compatible. More specifically, \cite{selfish-mining} showed that an adversarial miner, who possesses $\alpha$ fraction of the total hashrate in the network, can  achieve a revenue ratio $\rho$, that is more than its fair share $\alpha$. This is achieved by keeping the blocks it has mined private for a while according to the selfish mining algorithm and releasing them later to replace the blocks mined by honest miners, resulting in $\rho>\alpha$ in the long run. Moreover, the extra revenue ratio of selfish mining, \textit{i.e.}, $\rho-\alpha$, grows even further as the adversarial hash fraction in the system increases, which attracts the rational miners to join the adversary, eventually overtaking the chain completely. 

Later on, block withholding attacks and their effects on the revenue ratio have gained further attention and been studied extensively \cite{ optimal-selfish, stubborn-mining,on_the_sec_pow_gervais,sompolinsky2016bitcoin,prob-selfish-mdp-method,preneel_common_metrics,MDP_proof-systems-selfish, keller2024genericselfishminingmdp,yang2021-deep-dive-analysis-selfish-stubborn,optimal_stubborn_state_transition,optimal_stubborn_mdp,profitable_double_sp, Grunspan_profitable_double_sp,courtois2014subversiveminerstrategiesblock,miners_dilemma,fork_after_witholding_attack,power_adjusting,selfholding}. Sapirshtein et.~al.~\cite{optimal-selfish} use Markov Decision Processes (MDP) to provide $\epsilon$-optimal policies for selfish mining attacks that result in higher revenue ratio than the algorithm of \cite{selfish-mining}. In a concurrent work, Nayak et.~al.~\cite{stubborn-mining} provide an in-depth look at variations of selfish mining attacks, including the-now-famous stubborn mining  as well as its no-trail variations, and analyze their revenue ratio using Monte Carlo simulation techniques. Further, Gervais et.~al.~\cite{on_the_sec_pow_gervais} create an extended MDP problem where the model takes into account additional parameters such as the confirmation rule, network propagation dynamics, mining costs and eclipse attack. The authors of \cite{on_the_sec_pow_gervais} further modify the model to consider the profitability of the double-spending attack via the MDP and investigate the impacts of the parameters on the optimal policies. Similar to the study of \cite{on_the_sec_pow_gervais}, \cite{sompolinsky2016bitcoin} builds an MDP problem to investigate the profitability of the double-spending attack by slightly modifying the MDP model of \cite{optimal-selfish} that replaces the rewards of override blocks by double-spend blocks defined therein. The authors in \cite{sompolinsky2016bitcoin} further modify the MDP to investigate the strategies that optimally combine selfish mining attacks with double-spend attacks assuming a constant relation between the profits from the two attacks. The authors of \cite{prob-selfish-mdp-method} investigate a novel efficient approach to solve average reward ratio MDP of \cite{optimal-selfish} using probabilistic termination methods. Other generalizations of the MDP model also exist that analyze selfish mining in different proof systems and protocols \cite{preneel_common_metrics,MDP_proof-systems-selfish, keller2024genericselfishminingmdp,yang2021-deep-dive-analysis-selfish-stubborn,optimal_stubborn_state_transition,optimal_stubborn_mdp}. Selfish mining strategies can further be combined with block withholding attacks on pools to increase revenue ratio \cite{fork_after_witholding_attack,power_adjusting,selfholding}. Another line of research \cite{profitable_double_sp, Grunspan_profitable_double_sp} investigates the profitability of double-spending attacks using stochastic methods. Some of the recent works also focus on the profitability over time and use alternative metrics such as revenue per time or chain progress \cite{intermittent_mining,Grunspan_witholding_resilience,profit_lag,time_average_selfish_mining}.

An important feature of selfish mining attack model is the network influence, $\gamma$, of the adversary, which determines the fraction of honest miners who favor the adversarial chain in case of fork-ties.\footnote{However, in this model, also used in our paper, honest miners do not fork each other's blocks, which in turn implies that the model deals with the regime where the block arrival rate is slow as in the case of Bitcoin.} The parameter $\gamma$ provides a threshold for the profitability of selfish mining, \textit{i.e.}, the minimum adversarial hash fraction needed for the selfish mining attack to be profitable. In other words, $\gamma$ determines the minimum $\alpha$ needed such that $\rho(\alpha,\gamma)>\alpha$. The authors of \cite{selfish-mining} argue that the current Bitcoin protocol, which picks the first chain seen by a miner in case of a tie, is vulnerable as the cost of increasing network influence is minimal for the adversary. This in turn, resulted in search for  new consensus policies that increase the threshold and the security \cite{selfish-mining, freshness_preferred, Preventing_Selfish_Creation_Time, Publish_or_Perish,preneel_common_metrics}. Other studies provide methods to effectively calculate network influence $\gamma$ under various network models \cite{Gobel_selfish_mine_prop_delay, Impact_of_Network_Connectivity_on_Consensus}. We refer the interested reader to the survey in \cite{survey_double_sp_selfish} for a further read on selfish mining and double-spending attacks and related defense mechanisms.

\subsection{Our Contributions}
Double-spending attacks in blockchain literature usually assume an adversary that does not focus on coinbase rewards and purely tries to undermine the consistency of the protocol. Such an analysis that ignores the adversarial revenue loss is usually not realistic. Hence, in this paper, for the first time in the literature, we provide a rigorous and explicit stochastic analysis of the combination of the two celebrated attacks: double-spending and selfish mining. Following the attack prescription we provide, gives the adversary repeated chances at double-spending without incurring revenue losses. To do so, we define a selfish mining attack strategy, called $L$-stubborn mining, that not only tracks the length difference between the adversary's private chain and the public chain but also its length. Tracking the length of the private chain allows us to determine whether the attack can result in double-spending or not in addition to the revenue stolen from the honest miners. 

Essentially, in $L$-stubborn mining, the adversary keeps mining on its chain privately until it reaches length $L$ as long as it does not fall behind of the public honest chain. From this point onwards, the adversary does not risk falling behind the public honest chain, hence releases its private chain if and when the honest chain is one block behind.
Note that, parametrized by $L$, $L$-stubborn mining is a generalization of existing mining strategies in the literature, such as honest mining ($L=1$), selfish mining ($L=2$), equal fork stubborn mining ($L=\infty$) where $L$ represents the stubbornness level of the adversary. Thus, an optimization of $1\leq L\leq\infty$ results in optimal revenue ratios $\rho_{L^*}$ among the no-trail stubborn strategies.

More specifically, for each $\alpha$ and $\gamma$, we find the optimal $L^{*}$, \textit{i.e.}, the best strategy, under the condition that the adversary does not do any trailing behind, \textit{i.e.}, it accepts the honest branch whenever it falls behind. Similarly, we find the largest $L$, which we call $\Bar{L}$, such that $\rho>\alpha$, \textit{i.e.}, maximum stubbornness that still results in higher revenue ratio than honest mining strategy. Next, using the $k$-confirmation rule as a heuristic to reach consensus as is the convention, we argue that if $\Bar{L}> k$, then double-spending comes at no-cost. Hence, the results not only give a new understanding of selfish and stubborn mining but also provide parameter regimes of $\alpha$ and $\gamma$ where transactions are at risk from the perspective of double-spending. It is already known that when $\alpha > 0.5$ no transaction is safe. With the results we provide, for example for Bitcoin which follows $k$-confirmation rule with $k=6$, even if $\gamma=0$, every transaction is at risk of double-spend for $\alpha>0.409$. This, however, does not mean, that the transaction is going to be replaced with a conflicting one without fail, instead, it means, that trying the double-spending attack with no-trailing comes at no-cost to the adversary.

We also provide a modified version of $L$-stubborn mining attack, which we call $S$-stealth mining, that is more similar to the classical private attack, in the sense that it does not expose itself by prematurely publishing a competing adversarial block at the public honest chain’s tip and increases the success probability of double-spending attack at the cost of revenue. We analyze the properties of $S$-stealth mining as we do for the $L$-stubborn mining attack.

Our analysis considers average reward ratio MDP of the \cite{optimal-selfish,prob-selfish-mdp-method} for no-trail stubbornness and we provide a novel algorithm to solve the problem, which outputs the optimal values in finite steps. Our approach can be a basis to solve the general MDP problem, \textit{i.e.}, with trail stubbornness, directly, which we leave for future research. We summarize the contributions of our paper as follows:
\begin{itemize}
    \item Given the adversarial hash fraction $\alpha$ and adversarial network influence $\gamma$, we find the optimal level of no-trail stubbornness $L$ and provide the associated revenue ratio.
    \item We connect the attack to double-spending, as $L>k$ implies double-spending. Thus, given $\alpha$ and $\gamma$, we provide values of $k$ where $k$-confirmation rule is at risk of double-spending (which comes at no-cost to the adversary).
    \item We provide the double-spending probability, \textit{i.e.}, the risk, for each attack cycle.
    \item We provide a modified version of the attack that works by stealth in order to prevent exposure and increase the double-spending probability at the cost of mining revenue.
    \item We provide the minimum double-spend value needed for an attack to be profitable in the regimes where the scheme is less profitable than honest mining. 
\end{itemize}

Different from \cite{on_the_sec_pow_gervais, sompolinsky2016bitcoin}, we provide a simple formula to calculate revenue ratios, double-spend risks and optimal attacks. As we do not deal with MDPs, the numerical results in Section~\ref{sec::numerical} are extremely easy-to-analyze and provide a great insight at first sight. Moreover, the strategy we provide is easy-to-follow compared to complicated prescriptions of MDPs. We numerically compare $\rho_{L^*}$ with the revenue ratio of the $\epsilon$-optimal selfish mining attack to argue that our strategy performs close to the optimal solution of the MDP of \cite{optimal-selfish}. Hence, our results can be a substitute to rigorously analyze optimal mining revenue per time, a recent focus of research topic, instead of going through the tedious MDP and Monte Carlo runs  \cite{intermittent_mining,Grunspan_witholding_resilience,profit_lag,time_average_selfish_mining}.

In our analysis, we resort to the Bertrand's ballot problem\cite{ballot-problem}, Bailey's number\cite{baileys_number} and Catalan numbers\cite{catalan-numbers-book}, which have appeared explicitly in the blockchain literature in \cite{blockchains_biased_ballot_problem} where Chen et.~al.~calculate the probability that a chain wins a fork race given different mining probabilities depending on which fork is leading. Although Chen et.~al.~correctly argue that celebrated blockchain problems such as Nakamoto's security problem \cite{btc-whitepaper} and selfish mining phenomenon \cite{selfish-mining} can be cast as modified ballot problems, they only provide the probabilities of each fork being ahead and their asymptotic values. Catalan numbers further appeared in \cite{catalan-stubborn-grunspan,grunspan2020mathematics} where Grunspan and Marco make use of them to calculate the revenue ratio of the stubborn mining \cite{stubborn-mining} in closed form and in \cite{Goffard_Fraud_risk} where Goffard refines the mathematical model of double-spending.

Our paper is structured as follows. After defining the system model in Section~\ref{sec::model}, we provide the attackers' mining strategies and relevant definitions in Section~\ref{sec::strat_defn}. In Section~\ref{sec::analysis_l}, we rigorously analyze the revenue ratio of the $L$-stubborn attack as well as the double-spending that comes with it. We do the same for $S$-stealth mining strategy in Section~\ref{sec::stealth}. In Section~\ref{sec::numerical}, we provide the relevant numerical results. Finally, in Section~\ref{sec::conclusion}, we discuss the severity and potential implications of the attack as well as possible extensions.

\section{System Model}\label{sec::model}
We assume a blockchain running under the Nakamoto consensus, \textit{i.e.}, PoW with longest chain protocol. The mining model we consider is the same as the model used in the seminal work on selfish mining by Eyal and Sirer\cite{selfish-mining} and the subsequent works on the topic \cite{optimal-selfish, stubborn-mining,on_the_sec_pow_gervais,sompolinsky2016bitcoin,prob-selfish-mdp-method,preneel_common_metrics,MDP_proof-systems-selfish, keller2024genericselfishminingmdp,yang2021-deep-dive-analysis-selfish-stubborn,optimal_stubborn_state_transition,optimal_stubborn_mdp,profitable_double_sp, Grunspan_profitable_double_sp,courtois2014subversiveminerstrategiesblock,miners_dilemma,fork_after_witholding_attack,power_adjusting,selfholding}. More specifically, we assume that there are two groups of miners. The first group can be considered as a single-entity which constitutes $\alpha=1-\beta<0.5$ fraction of the mining power in the system and acts adversarially, \textit{i.e.}, does not follow the longest-chain protocol and can decide to mine privately without sharing the blocks it has mined. In this paper, we call this entity as \textit{the adversary}. The second group comprises the honest miners who follow the longest-chain protocol, \textit{i.e.}, mine on top of the longest-chain they have seen and release the newly mined blocks immediately. In case of a tie, the honest miners prefer the chain they first became aware of.

In this paper, we only consider the blocks that contain a valid PoW (Proof-of-Work) and call the blocks mined by an honest miner as an honest block whereas a block with a valid PoW mined by the adversary is called an adversarial block. Here, we note that the inter-arrival times between blocks follow an exponential distribution due to the properties of PoW mining. When a block is mined by an honest miner, it is subject to a small propagation delay. During this propagation delay, if the adversary has a competing block at the same height that is kept private, it can rush to release the block. In this context, \textit{releasing a block} means that the adversary shares the block with the rest of the network, \textit{i.e.}, the honest miners. This action is called \textit{matching}, and subsequently, $\gamma$ fraction of the honest miners receive the adversarial block first and mine on top of it, whereas the rest of the honest miners receive the honest block first and mine on top of it. Hence, $\gamma$ is called the \textit{adversarial network influence} or \textit{rushing factor}. 

However, during the propagation delay of an honest block, it is inherently assumed that no new blocks are mined by honest miners or the adversary. Hence, all honest blocks as well as the blocks released by the adversary are treated as public blocks in this model, \textit{i.e.}, when an honest miner mines a block, all other honest miners also become aware of the block immediately although the adversary can rush and release a block at the same height to influence $\gamma$ fraction of the honest miners who then prefer the adversarial branch. As a result, in this model, honest miners do not fork each other's blocks, which in turn implies that the model deals with the regime where the block arrival rate is slow as in the case of Bitcoin. We relegate a detailed discussion on the matter to Section~\ref{sec::prop_delays}.

Here, we assume the adversary is aware of any block in the network as soon as that block is mined. Any block that is known/observed by honest miners is called \textit{public block} and the \textit{public chain} refers to the longest chain that exclusively consists of public blocks. However, during a fork where the adversary matches an honest block on the tip of the longest chain by releasing its private block at the same height, for the sake of distinguishing the two forks that are public and of equal length, we call the branch released by the adversary as the adversarial public branch.

The blockchain protocol updates the puzzle difficulty threshold so that the average inter-arrival time between blocks is constant. Each block rewards its miner a predetermined amount of coins, \textit{i.e.}, coinbase, which we assume is constant for the foreseeable future. If each miner acts honestly, the coinbase reward they get is proportional to their mining power in the network in the limit. However, adversarial strategies which deviate from the longest-chain protocol result in rewards disproportional to the mining powers. We refer to the adversarial strategies that aim to increase their fraction of the coinbase rewards as selfish mining strategies. Although these selfish mining strategies initially decrease the rewards, they can be seen as an attack on the block interval adjustment \cite{grunspan2019-profitability-selfish-mining}, which in turn increases the rewards after the difficulty is adjusted.

In this work, we use the heuristic of $k$-confirmation rule for reaching consensus. A block in this system is considered to be confirmed according to a $k$-deep confirmation rule if it is part of a longest chain available to the honest miners. Instead of defining and using conflicting $\mathsf{tx}$s to refer to double-spending events, in this paper, we simply use the notion of multiple blocks at the same height being confirmed (at the same time or at two different points in time) as double-spending events. The heuristic of $k$-confirmation rule for reaching consensus is the convention in the literature and part of our system model where we pick a fixed $k$ to analyze the double-spending event. This is because honest miners are assumed to not afford waiting too long to confirm a block. However, it is important to mention that if latency to confirm a block is not of importance to honest miners, they could avoid the poor operating procedure of double-spending by waiting sufficiently long before confirming a block.

\subsection{Propagation Delays}\label{sec::prop_delays}

From the perspective of the propagation delays, this model has been used in the majority of the selfish mining literature and is still the convention \cite{optimal-selfish, stubborn-mining,on_the_sec_pow_gervais,sompolinsky2016bitcoin,prob-selfish-mdp-method,preneel_common_metrics,MDP_proof-systems-selfish, keller2024genericselfishminingmdp,yang2021-deep-dive-analysis-selfish-stubborn,optimal_stubborn_state_transition,optimal_stubborn_mdp,profitable_double_sp, Grunspan_profitable_double_sp,courtois2014subversiveminerstrategiesblock,miners_dilemma,fork_after_witholding_attack,power_adjusting,selfholding,catalan-stubborn-grunspan,profit_lag,werlman,Deep_Bribe,squirrl,intermittent_mining,Grunspan_witholding_resilience,time_average_selfish_mining}. Essentially, if blocks can be mined during propagation delays, the analysis has to keep track of multiple (potentially infinite) different honest miner views that evolve as time goes by. Hence, in the selfish mining literature, the assumption of no block mining during propagation delays has been adopted for tractability purposes. On the other hand, studies on double-spending analysis, such as \cite{nakamoto-always-wins,cao2023tradeoff,our-sec-lat-extended}, consider adversarial blocks mined during the propagation delays, because from the perspective of the consistency guarantees, worst-case scenarios are relevant. Hence, these studies usually assume that the adversary controls the propagation delays as long as each propagation delay is lower than a maximum $\Delta$ and ties are broken in its favor. As a result, these analyses usually focus on a single worst-case honest miner view \cite[Section~III.A, \textit{weakest possible view}]{cao2023tradeoff}. However, such a situation is extremely pessimistic as our goal is not to give worst-case analysis of revenue ratios in selfish mining, rather a realistic evidence that the adversary is incentivized to deviate from honest protocol. We also note that, optimal selfish mining analysis has been done previously in such a pessimistic setting \cite[Lemma~4]{our-queue-sec-ext-version}. The results therein follow an exponentially bounded delay structure, however, they are easily extendable to bounded $\Delta$ delay model of \cite{nakamoto-always-wins}, \textit{i.e.}, by replacing $P'$ matrix of \cite{our-queue-sec-ext-version} with $P$ matrix of \cite{our-sec-lat-extended} and redoing the subsequent analysis of \cite[Lemma~4]{our-queue-sec-ext-version}.

Further, under Nakamoto consensus, given the block creation rate $\lambda$, if $\lambda\Delta$ is not small, the consistency of the system is at threat and selfish mining is a less relevant concern. In such systems, the research focuses on different consensus models such as \cite{Sompolinsky2015SecureHT,PHANTOM_GHOSTDAG,SPECTRE}. On the other hand, as long as $\lambda\Delta$ is small, it can be shown that the probability of more than $2$ blocks being mined within a $\Delta$ interval is negligible \cite{Impact_of_Temporary_Fork,Impact_of_Network_Connectivity_on_Consensus,sakurai2024modelbasedanalysisminingfairness}. Thus, in the realistic scenario of one block being mined during the propagation delay of a block, during forks, the honest network divides between two branches. As a result, in the long run, the parameter $\gamma$ used in selfish mining literature is still a relevant and reasonable assumption. Further, in \cite[Section~V]{Impact_of_Temporary_Fork} the authors give a formula to estimate $\gamma$ given propagation delays between mining entities and validate the estimation with network simulations. 

Essentially, as things become immediately untractable when one considers multiple miners' views during network delays, the system model with $\alpha$ and $\gamma$ serve as tractable and efficient parameters to model selfish mining attacks in a simplified manner that still roughly translates to a system with propagation delays where $\gamma$ is the main parameter of interest from the perspective of the delays and fork conflicts and should be estimated based on the observed delays. Here, note that, propagation delays also affect the apparent hashrate of a mining entity. In other words, the apparent hashrate might deviate from the fraction of computational power the entity possesses due to the propagation delays and network structure even if everyone follows honest mining strategy. Notice that, propagation delays result in forks and the block released and observed first by other miners does not always win the fork race. Hence, one can also think of $\alpha$ as the apparent hash rate instead of the fraction of computational power. For example, studies such as \cite{sakurai2024modelbasedanalysisminingfairness} consider different propagation delays between different miners to evaluate fairness, which is roughly defined as the difference between the fraction of computational power and the apparent hashrate a mining entity has. We believe that modeling such a system to estimate $\alpha$ and $\gamma$ values used in this paper are possible however out of scope of this paper. For further details about the propagation delays and their impact on fork races and fairness, we refer the reader to \cite{Impact_of_Temporary_Fork,Impact_of_Network_Connectivity_on_Consensus,sakurai2024modelbasedanalysisminingfairness}.

\section{Mining Strategies and Definitions}\label{sec::strat_defn}
\subsection{$L$-Stubborn Mining}\label{sec::l-stub}
The selfish mining strategies are extensively analyzed in the literature with notable works such as \cite{selfish-mining,stubborn-mining,optimal-selfish}. In this work, we consider a sub-group of strategies from all possible states and action space, which we call $L$-stubborn\footnote{In this paper, $L$-stubborn mining is different than the lead-stubbornness (L-stubbornness) defined in \cite{stubborn-mining}. There, ``L'' stands for the word ``lead.'' Here, $L$ refers to an integer which is the length of the adversarial chain with respect to a common ancestor block between the honest and the adversarial chains.} mining. Here, we explain how $L$-stubborn mining strategy works explicitly and later in Appendix~\ref{app::state_action_stub_stealth}, we present a more formal description of state and action spaces for the same strategy.

To describe the strategy, let us assume that all participants start with the same chain, which we call the offset chain. For the sake of simplicity, we denote the height of the offset chain as zero.  Both the adversary and the honest miners mine on top of the offset chain. We denote the highest adversarial block height as $A$ and the chain starting from the offset and ending in the adversarial block at height $A$ as the $A$-chain. We denote the highest honest block height as $H$ and the chain starting from the offset and ending in the honest block at height $H$ as the $H$-chain.

Starting from the offset chain:
\begin{enumerate}
    \item If the honest miners are the first to mine, then the adversary accepts the new block and redefines the offset chain with the new block on its tip.
    \item If the adversary is the first to mine a block on top of the offset chain, \textit{i.e.}, at height $1$, it keeps the block private and  continues to mine privately on top of its block.  At any time later, if the honest miners mine a block at height $h\leq A<L$, the adversary releases its block at the same height $h$ immediately which we call \textit{matching}. In accordance with the network model defined earlier, $\gamma$ proportion of the honest miners accept and mine on top of the adversarial block instead of the honest block.
    \begin{enumerate}
        \item If honest miners manage to mine a block on a height before the adversary before $A=L$, \textit{i.e.}, $L \geq H=A+1$, the adversary accepts the $H$-chain and redefines the offset chain. Such an action of accepting $H$-chain and abandoning $A$-chain is called \textit{adopt}.
        \item If honest miners do not manage to mine a block on a height before the adversary and $A$ chain reaches the length $A=L$, then the adversary keeps mining privately until $A=H+1$, at which point, it releases the $A$-chain and redefines the offset chain. Such an action of releasing an $A$-chain with $A>H$ is called \textit{override}.
    \end{enumerate}
\end{enumerate}
An attack cycle is a process that starts when an offset chain is defined, and ends when a new offset chain is defined. Note that, offset chain is redefined whenever an \textit{adopt} or \textit{override} action (event) happens, which we call an \textit{offset event} in general. If during an attack cycle, the length of the $A$-chain reaches $L$, \textit{i.e.}, $A=L$, then we call the attack successful. On the other hand, if the length of the $H$-chain surpasses the length of the $A$-chain before $A$ reaches $L$, we call the attack cycle unsuccessful.

Here, we consider an example that is critical in understanding the model and the subsequent analysis in later parts of the paper. Let us assume $L=7$ and currently $A=3$, $H=0$ at an attack cycle where the private $A$-chain is $A=\mathsf{B'}|\mathsf{B_2'}|\mathsf{B_3'}$ where `$|$' represents two consecutive blocks chained/concatenated. Assume the next block mined is an honest block $\mathsf{B}$ which is at the same height as $\mathsf{B'}$ since honest miners are not aware of any block at that height. As soon as $\mathsf{B}$ is mined, the adversary matches it, \textit{i.e.}, releases $\mathsf{B'}$. At this point, $A=3$, $H=1$ and $\gamma$ fraction of honest miners are mining on top of $\mathsf{B'}$ whereas the rest of the honest miners mine on top of $\mathsf{B}$. If the next block is honest, \textit{i.e.}, $\mathsf{B_2}$, then $A=3$, $H=2$ and the adversary matches $\mathsf{B_2}$ with $\mathsf{B_2'}$ but there are two cases regarding the prefix of $\mathsf{B_2}$. Either $H=\mathsf{B}|\mathsf{B_2}$ with probability (w.p.) $1-\gamma$ or $H=\mathsf{B'}|\mathsf{B_2}$ w.p. $\gamma$. Hence, during an attack cycle, the prefix of $H$-chain can change and contain some number of adversarial blocks.\footnote{A natural question is, if $A=\mathsf{B'}|\mathsf{B_2'}|\mathsf{B_3'}$ and $H=\mathsf{B'}|\mathsf{B_2}$, then why not redefine offset chain with $\mathsf{B'}$ at its tip since it is accepted by everyone. If we do so, then $A=2$ with blocks $A=\mathsf{B_2'}|\mathsf{B_3'}$ and $H=1$ with block $\mathsf{B_2}$. However, this renders attack cycles as dependent and we lose track of the double-spending between blocks $\mathsf{B}$ and $\mathsf{B'}$. This is an essential design choice for tractability purposes, see Section~\ref{sec::extensions} for possible extensions.}

\subsubsection{An Example: $4$-Stubborn Mining}
\begin{figure}[t]
\centerline{\includegraphics[width=\columnwidth]{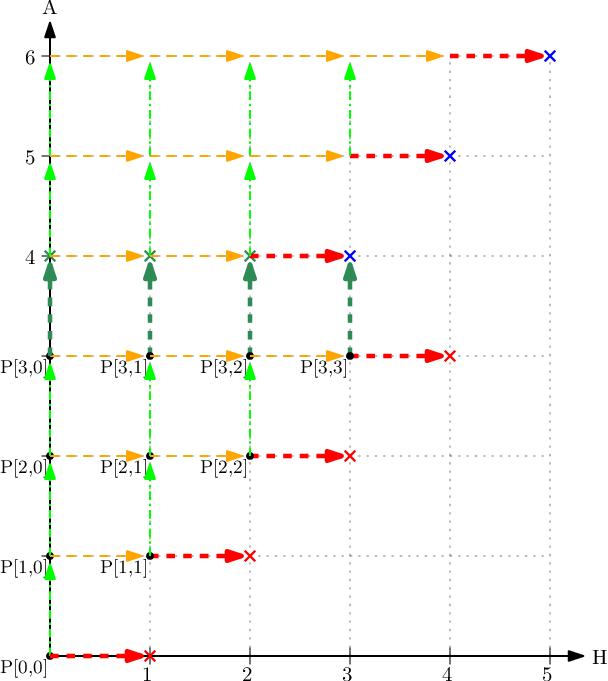}}
	\caption{Sample paths of $4$-stubborn mining.}
	\label{fig::coordinate}
\end{figure}

In \figref{fig::coordinate}, we display the sample paths of $4$-stubborn mining, \textit{i.e.}, $L=4$. The  $y$-axis and $x$-axis represent $A$ and $H$, \textit{i.e.}, the length of $A$-chain and $H$-chain, respectively. An attack starts from point $(0,0)$ and arrivals of adversarial (honest resp.) blocks are represented with light-green (orange resp.) arrows. A red arrow shows that the arrival of honest block has ended the cycle. A blue (red resp.) cross represents the endpoint of a successful (unsuccessful resp.) attack cycle. A dark-green arrow represents the arrival of an adversarial block that guarantees a successful attack cycle. The dark-green crosses represent the first time cycles are guaranteed to be successful and not ended yet. The cycles that pass through the dark-green crosses eventually end with an honest arrival in blue cross points $(x,x-1)$ where $x>4$, \textit{i.e.}, when the difference of chain lengths reduces to $1$. The notion of $P[n,m]$ at $(A,H)=(n,m)$ displayed in \figref{fig::coordinate} represents the number of distinct paths an attack cycle can take to reach the point $(A,H)=(n,m)$. In Appendix~\ref{sec::preliminaries} we give the formula for $P[n,m]$ as well as explain the reasoning behind.

Note that there is a third dimension of \figref{fig::coordinate}, which tracks the number of adversarial blocks in the prefix of $H$-chain. Further, the dynamics of this dimension is Markovian since each honest arrival either increases the previous value to $H-1$ (w.p. $\gamma$) or stays the same (w.p. $1-\gamma$) and does not depend on anything else. This Markovian structure implies that we can separate the third dimension and treat it independently which simplifies our analysis, hence, we do not plot it; see Appendix~\ref{app::state_action_stub_stealth} for more details.

\subsubsection{Relation to the Existing Works}
For $L=1$ the strategy reduces to honest mining strategy, whereas for $L=2$ the strategy reduces to the celebrated selfish mining attack of \cite{selfish-mining}. On the other hand, for $L=\infty$ the strategy becomes the equal fork stubborn mining strategy described in \cite{stubborn-mining} and analyzed in closed-form with Catalan numbers in \cite{catalan-stubborn-grunspan}. In this paper, we are interested in giving a method for calculating the revenue ratios for any given $L$ as well as deriving a method to find the best possible $L$ given $\alpha$ and $\gamma$. The resulting strategy does not allow any trail stubbornness discussed in \cite{stubborn-mining}, \textit{i.e.}, the adversary accepts the honest chain whenever it falls behind. Hence, we can only claim the optimality of the strategy under action spaces with independent attack cycles that allow no-trail stubbornness. We argue that the parameter $L$ represents the level of stubbornness in selfish mining and our analysis provides the optimal level of stubbornness as well as the maximum stubbornness such that the attack is still more profitable than honest mining.

In \cite{optimal-selfish}, the authors use an undiscounted average reward MDP to provide $\epsilon$-optimal policies, i.e., Optimal Selfish Mining (OSM) MDP. The authors of \cite{optimal-selfish} further argue about combining double-spending attack with selfish mining attack, however, they do not provide any concrete method or analysis. Later, in \cite{sompolinsky2016bitcoin}, the same authors provide an MDP that accounts for rewards of double-spending attack. However, the results and strategies provided in \cite{sompolinsky2016bitcoin, optimal-selfish, on_the_sec_pow_gervais} are complicated with no easy-to-follow structure. On the other hand, $L$-stubborn mining that we consider in this paper is a policy which results in higher mining rewards than honest mining and we provide an efficient algorithm to obtain the optimal block mining revenue through optimizing $L$, which has strong implications on double-spending strategies with a clear and easy-to-follow analysis under $k$-confirmation rule, which we explain next.

\subsection{Double-Spend Events}
Here, we rigorously define the events related to a double-spending attack.
\begin{definition}
    {\normalfont \textbf{($\boldsymbol{k}$-Confirmation rule)}} A block is said to be confirmed according to the $k$-confirmation rule if it is part of a longest public chain and at least $k$-deep, \textit{i.e.}, there are at least $k-1$ public blocks mined on top of it in this longest chain.
\end{definition}
\begin{definition}
    {\normalfont \textbf{(Double-Spend Events)} Consider an honest block $\mathsf{B}$ and an adversarial block $\mathsf{B'}$ at the same height. We consider three disjoint events regarding the confirmation of $\mathsf{B}$ and $\mathsf{B'}$:
    \begin{enumerate}
        \item {\textbf{Service}} If $\mathsf{B}$ is confirmed and makes it into the offset chain, we call the event \textit{Service}.
        \item {\textbf{Move-Funds}} If $\mathsf{B}$ is never confirmed, we call the event \textit{Move-Funds}.
        \item {\textbf{Double-Spending}} If $\mathsf{B}$ is confirmed but does not make it into the offset chain, we call the event \textit{Double-Spending}.\footnote{The adversarial block $\mathsf{B'}$ is assumed to contain a transaction where the adversary simply moves some funds between its sybil accounts, whereas the honest block $\mathsf{B}$ is assumed to contain a conflicting transaction where the adversary pays a merchant for some service using the same funds. As observed by \cite{optimal-selfish}, Service or Move-Funds events should not be seen as a loss from the perspective of the adversary.
        
        It is worth mentioning that according to the adversarial action space we consider in this paper, the occurrence of the event \textit{Service} implies that $\mathsf{B'}$ is never confirmed, \textit{i.e.}, only $\mathsf{B}$ is confirmed, although an unrestricted state/action space does not exclude such an instance. Similarly, \textit{Move-Funds} implies that only $\mathsf{B'}$ is confirmed and \textit{Double-Spending} implies that first $\mathsf{B}$ is confirmed but then $\mathsf{B'}$ is confirmed and makes into the offset chain. These three events exhaust all the possibilities regarding the confirmation of block $\mathsf{B}$ and $\mathsf{B'}$ according to the attacks we consider in this paper.}
    \end{enumerate}}
\end{definition}

Throughout the paper, for each attack cycle, we consider the first block of the $H$-chain after the offset event, $\mathsf{B}$, as the target honest block for double-spending that is to be attacked. Notice the relation between $L$-stubborn mining and $k$-confirmation rule. If $(k+1)$-stubborn mining attack is more profitable than honest mining, at every attack cycle, the adversary gets a shot at double-spending at no-cost on the first honest block while implementing the $(k+1)$-stubborn mining strategy.

Before the offset event happens, if $A>k$, this implies that at some point $H=k$ and the adversary will release $A$-chain later on. Depending on whether $H$-chain contains $\mathsf{B'}$ or $\mathsf{B}$, either Move-Funds or Double-Spending happens. On the other hand, before the offset event happens, if $A<H\leq k$, depending on whether $H$-chain contains $\mathsf{B'}$ or $\mathsf{B}$, either Move-Funds or Service happens. If an attack cycle ends with Move-Funds event, later, the adversary can try double-spending again using the same funds. One could also rescale the probabilities of Double-Spending and Service by excluding Move-Funds event from the perspective of the actual funds associated with $\mathsf{B}$ and $\mathsf{B'}$ since eventually the fund is going to result in Service or Double-Spending. We omit the rescaling in our calculations and numerical results, as rescaling is straightforward to do and we would like to present the exact probability  of Double-Spending event in each attack cycle.

We also note that if we consider transactions instead of blocks, then the time $\mathsf{tx}$ is released in the network is another important issue while considering double-spending attacks. In general, it can be shown that a double-spending attack can be successful w.p. $1$ if the adversary is allowed to pick the time $\mathsf{tx}$ is released, since it can build an arbitrarily high pre-mining gain \cite{sompolinsky2016bitcoin}. The pre-mining gain refers to the difference between the private chain of the adversary and the public honest chain, in general \cite{sompolinsky2016bitcoin}. The approach we take here is more conservative since we only consider multiple block confirmations at the same height as double-spending, hence the pre-mining gain is $0$.

\subsection{$S$-Stealth Mining}\label{sec::s-steal}
Notice the difference between Nakamoto private attack and $(k+1)$-stubborn mining. The first major difference is $(k+1)$-stubborn mining gives up the attack when $A$-chain falls behind $H$-chain, \textit{i.e.}, no-trailing behind is allowed. Another major difference is, during a cycle, $(k+1)$-stubborn mining matches each honest block with an adversarial one which can expose the attack as two chains with conflicting blocks are competing openly in the network, arousing the merchant's suspicion. The matching also decreases Double-Spending probability while increasing Move-Funds probability, which we show later.

We address some of these issues by modifying $L$-stubborn mining in a way that only allows matching if both double-spending and the risk of losing mining revenue are imminent. More specifically, in $S$-stealth mining, at every attack cycle the adversary mines privately on top of $A$-chain until it reaches the length $A=S$. If the $A$-chain falls behind the $H$-chain before reaching the length $S$, it accepts the $H$-chain and redefines the offset chain to start a new attack cycle. In $S$-stealth mining, the adversary takes the matching action only in the following case: $A=S-1$ and $H=S-2$ and the next block is an honest block. If the $A$-chain reaches length $S$, the adversary waits until $A=H+1$, at which point it releases the $A$-chain to override. 

Thus, $S$-stealth mining strategy, which is investigated in Section~\ref{sec::stealth}, is essentially the same strategy as $L$-stubborn mining strategy except that the matching action only happens when $H=A=S-1$. We note that $S$-stealth mining does not allow any trailing behind either, hence, there is still a difference between a private attack and $S$-stealth mining. Note that, similar to $L$ of $L$-stubborn mining strategy, we can view $S$ of $S$-stealth mining strategy as a parameter of stubbornness. We present a more formal description of state and action spaces for $S$-stealth mining in Appendix~\ref{app::state_action_stub_stealth}.

\subsection{Definitions and Notations}
In this section, we give definitions and notations that are used frequently throughout the paper. For the convenience of the reader, some definitions provided in the other parts of the paper are listed here as well.

\begin{definition} {\normalfont \textbf{(Model parameters)}} The fraction of the total hashrate the adversary possesses is denoted as $\boldsymbol{\alpha<0.5}$. The remaining miners are considered to be honest and possess $\boldsymbol{\beta=1-\alpha}$ fraction of the hashrate. When a public fork exists, $\boldsymbol{\gamma}$ fraction of the honest miners prefer the adversarial branch.
\end{definition}

\begin{definition} {\normalfont \textbf{(Mining strategies)}} The following strategies are frequently mentioned in the paper:
\begin{enumerate}
    \item \textbf{$\boldsymbol{L}$-stubborn mining} strategy refers to the mining strategy for which the procedure is explicitly explained in Section~\ref{sec::l-stub} and formally defined in Appendix~\ref{app::state_action_stub_stealth}.
    \item \textbf{Honest mining} strategy refers to the mining strategy when the adversary strictly follows the honest protocol and releases a block as soon as it is mined. The strategy can be explicitly obtained by plugging $\boldsymbol{L=1}$ into the procedure  explained in Section~\ref{sec::l-stub}.
    \item \textbf{Selfish mining} strategy refers to the mining strategy proposed by Eyal and Sirer \cite{selfish-mining}. The strategy can be explicitly obtained by plugging $\boldsymbol{L=2}$ into the procedure  explained in Section~\ref{sec::l-stub}. We also give an explicit summary of the attack in Appendix~\ref{sec::app-existing} as well as how $L$-stubborn mining generalizes the attack.
    \item \textbf{Equal fork stubborn mining} strategy refers to the mining strategy proposed by \cite{stubborn-mining}. The strategy can be explicitly obtained by plugging $\boldsymbol{L=\infty}$ into the procedure  explained in Section~\ref{sec::l-stub}.
    \item \textbf{$\boldsymbol{S}$-stealth mining} strategy refers to the mining strategy for which the procedure is explicitly explained in Section~\ref{sec::s-steal} and formally defined in Appendix~\ref{app::state_action_stub_stealth}.
\end{enumerate}

\end{definition}

\begin{definition}\label{def::rev_ratios}
    {\normalfont \textbf{(Revenue ratio)}} Revenue ratio of the adversary refers to the fraction of the blocks created by the adversary in the longest chain in the long run. In other words, let $N_{A}^t$ and $N_{H}^t$ denote the the number of adversarial and honest blocks in the longest chain at time $t$ respectively. Then,
    \begin{align}
        \rho=\lim_{t \to \infty} \frac{N_{A}^t}{N_{A}^t + N_{H}^t}.
    \end{align}
    Here, \begin{enumerate}
        \item $\rho_L$ is the revenue ratio under the $L$-stubborn mining strategy, for which an explicit equation is provided in Theorem~\ref{thm::revenue}, equation \eqref{eq::revenue}. Here:
        \begin{enumerate}
            \item $\rho_1$ is the revenue ratio when the adversary employs the honest mining strategy, where
        \begin{align}
            \rho_1=\alpha.
        \end{align}
        \item $\rho_2$ is the revenue ratio of selfish mining strategy, where
        \begin{align}
            \rho_2=\frac{\alpha\beta^2(4\alpha+\gamma(1-2\alpha))-\alpha^3}{1-\alpha(1+(2-\alpha)\alpha)}.\label{eq::self_formula}
        \end{align}
        \item $\rho_\infty$ is the revenue ratio of equal fork stubborn strategy, where
        \begin{align}
            \rho_{\infty}&=\frac{\alpha}{\beta}-\frac{(1-2\alpha)(1-\gamma)}{\beta\gamma}(1-\beta C((1-\gamma)\alpha\beta))
        \end{align}
        and $C(x)$ is given in \eqref{eq::catalan_generator}.
        \end{enumerate}
        \item $\sigma_S$ is the revenue ratio of $S$-stealth mining strategy, for which an explicit equation is provided in Theorem~\ref{thm::revenue_stealth}, equation \eqref{eq::stealth_revenue}.
    \end{enumerate}
    \begin{remark}
        In the long run, for every coin issued by the protocol, the adversary gets $\rho$ fraction of it. Note that, for a given $(\alpha,\gamma)$ pair, if a specific mining strategy by the adversary results in a revenue ratio higher than the revenue ratio of honest mining strategy, \textit{i.e.}, $\rho>\alpha$, the system is not incentive compatible for the $(\alpha,\gamma)$ pair. 
    \end{remark}
    \begin{remark}
        Some mining strategies potentially result in double-spending profits in addition to the revenues from the coinbase rewards of the blocks. These additional profits are not considered in revenue ratio as it is not the convention in selfish mining literature. Instead, we define a \textbf{combined revenue ratio} in Section~\ref{sec::comb_revenue} (and in Section~\ref{sec::comb_revenue_stealth}) that accounts for the additional profit coming from the double-spending similar to the \textit{subversion gain} defined in \cite{preneel_common_metrics}.
    \end{remark}
\end{definition}
\begin{remark}
    \textbf{Normalized revenue ratio} is defined as $\frac{\rho}{\alpha}$, which is a measure of how much the revenues are scaled when a miner employs a non-honest mining strategy compared to the honest mining strategy.
\end{remark}

\begin{definition}\label{def::l-values}
    $L^*$ refers to the value of $L$ that maximizes the revenue ratio of $L$-stubborn mining, \textit{i.e.}, 
    \begin{align}
    L^*&=\arg\max_{L} \rho_L.
    \end{align}
    Further, $\Bar{L}$ refers to the largest value of $L$ such that the revenue ratio of $L$-stubborn mining is higher than the honest mining strategy,
    \begin{align}
        \Bar{L}&=\sup\{L:\rho_L\geq\alpha\}.
    \end{align}
    Similarly, for $S$-stealth mining, we define
\begin{align}
    S^*&=\arg\max_{S} \sigma_S,\\
    \Bar{S}&=\sup\{S:\sigma_S\geq\alpha\}.
\end{align}
\end{definition}

We note that the following are defined in Section~\ref{sec::l-stub}: $L$, $A$, $H$, $A$-chain, $H$-chain, \textit{offset chain}, \textit{offset event}, \textit{matching}, \textit{override}, \textit{adopt} actions. As it is not possible to define them without completely going through the algorithm of $L$-stubborn mining, we refer the reader to Section~\ref{sec::l-stub} where they are defined.

\section{Analysis of $L$-Stubborn Mining}\label{sec::analysis_l}
We start by analysis of the revenue ratio of the adversary when it employs the $L$-stubborn mining strategy. Instead of building a complex Markov chain and going through the tedious analysis of steady-state distribution as it is done in \cite{selfish-mining}, we can analyze the distinct possible paths an attack cycle can take. 

\begin{lemma}\label{lemma:success}
    Under the $L$-stubborn mining strategy, the probability that $A$-chain reaches length $L$ during an attack cycle while the maximum height of the honest block is $H=m<L$, is
    \begin{align}
        P_s(L,m)=\frac{L-m}{L+m} {L+m \choose L}\alpha^{L}\beta^{m}.
    \end{align}
\end{lemma}

\begin{Proof}
     We distinguish the last adversarial block from the previous $L-1$ adversarial blocks to prevent double-counting. The number of distinct paths with $H=m<L$ blocks and $A=L$ blocks, where the last block is adversarial, is equal to $P[L-1,m]$; see \figref{fig::coordinate} for an example where $L=4$.
\end{Proof}

\begin{lemma}\label{lemma:unsuccess}
    The probability that an attack cycle is unsuccessful and ends with $H=n+1$ which has $i\leq n$ adversarial blocks in its prefix (after the offset chain) is
    \begin{align}
        P_u(n,i)=\frac{1}{n+1} {2n \choose n}\alpha^{n}\beta^{n+1}(1-\gamma)^{n-i}\gamma^{\mathbbm{1}_{i}}.
    \end{align}
\end{lemma}

\begin{Proof}
    The end of the unsuccessful attack cycle implies that the last block that arrives has to be honest and the adversary was never behind up until that the last block. As a result, the arrival sequence of honest and adversarial blocks creates a Dyck string with length $n<L$. There are $C_n$ number of such strings, where $C_n$ is the $n$-th Catalan number defined in \eqref{eq::catalan_formula} in Appendix~\ref{sec::preliminaries}. Together with the last honest block each arrival sequence has probability $\alpha^n \beta^{n+1}$. Hence, an attack cycle ends unsuccessfully with the honest block at height $n+1$ w.p. 
    \begin{align}
        P_u(n)=\frac{1}{n+1} {2n \choose n}\alpha^{n}\beta^{n+1}.
    \end{align}
    Further, each time an honest block arrives, it is either mined by the honest miners who are under the influence of adversarial view w.p. $\gamma$, or it is mined on top of the last honest block. As a result, no adversarial block is contained in the prefix of the honest block on height $n+1$ w.p. $(1-\gamma)^{n}$ or $i>0$ adversarial blocks are contained w.p. $(1-\gamma)^{n-i}\gamma$. Note that, $\mathbbm{1}_{i}=1$ if $i\neq 0$ and zero otherwise.
\end{Proof}

Next, we put Lemma~\ref{lemma:success} and Lemma~\ref{lemma:unsuccess} together to obtain the revenue ratio of selfish mining.

\begin{theorem}\label{thm::revenue} \textbf{(Revenue Ratio)}
    The revenue ratio of $L$-stubborn mining is given as
    \begin{align}
        \rho_L=\frac{L_s+L_{u,a}}{L_s+L_u}, \label{eq::revenue}
    \end{align}
    where
    \begin{align}
        L_s&=\sum_{m=0}^{L-1} P_s(L,m)\left(L+\frac{(L-m-1)\alpha}{1-2\alpha}\right),\label{eq::suc_a_blocks}\\
        L_{u,a}&=\sum_{n=0}^{L-1}\sum_{i=0}^{n} P_u(n,i) i, \label{eq::unsuc_a_blocks}\\
        L_{u}&=\sum_{n=0}^{L-1}P_u(n) (n+1).\label{eq::unsuc_t_blocks} 
    \end{align}
\end{theorem}

\begin{Proof}
    First note that, in the limit, the longest chain is essentially the offset chain that grows whenever an \textit{adopt} or \textit{override} happens. As a result, the revenue ratio of the attacker is the number of the blocks created by the attacker in the offset chain divided by the total length of the offset chain. As the \textit{offset event} is a stopping time, we can simply focus on a single attack cycle and calculate the expected number of blocks created by the adversary that become part of the offset chain denoted as $O_A$. Dividing this number to the expected growth of the offset chain in an attack cycle, denoted as $O_T$, gives us the revenue ratio. In other words, by strong law of large numbers, under the $L$-stubborn strategy, we have
    \begin{align}
        \rho_L=\lim_{t \to \infty} \frac{N_{A}^t}{N_{A}^t + N_{H}^t}=\frac{\mathbb{E}[O_A]}{\mathbb{E}[O_T]}.
    \end{align}
    In Lemma~\ref{lemma:success} and Lemma~\ref{lemma:unsuccess}, we provide the probability for each possible outcome of an attack cycle. If an attack cycle is successful, we have an $A$-chain at length $L$ and the maximum height of the honest block is $H=m<L$ at some time within the attack cycle. Clearly, if $L-m>1$, the adversary will continue the cycle until the difference goes down to $1$. Thus, given the attack is successful, using Wald's identity provided in Appendix~\ref{subsec::walds-identity}, the expected number of blocks that become part of the offset chain in this case is $L+\frac{(L-m-1)\alpha}{1-2\alpha}$ and all these blocks are adversarial. On the other hand, if the attack cycle is unsuccessful, $i$ adversarial blocks still make it into the offset chain w.p. $P_u(n,i)$ and the total number of blocks that make it into the offset chain is $n+1$. As a result, the expectation of the two cases are denoted as $L_{u,a}$ and $L_u$, respectively. Putting everything together, it is clear that $\mathbb{E}[O_A]=L_s+L_{u,a}$ and $\mathbb{E}[O_T]=L_s+L_{u}$, giving the desired result in \eqref{eq::revenue}.
\end{Proof}

\subsection{Optimizing the Revenue Ratio}
As we mentioned earlier, our $L$-stubborn mining strategy is a generalization and extension of the existing mining strategies of honest mining $(L=1)$, selfish mining \cite{selfish-mining} $(L=2)$, equal fork stubborn mining \cite{stubborn-mining} $(L=\infty)$. As a result, for a given $(\alpha,\gamma)$ pair, a natural question is, what integer value of $L$ maximizes the revenue ratio $\rho_L$. To find the optimal $L$, one could evaluate $\rho_L$ in \eqref{eq::revenue} and find the $L$ that provides the highest revenue ratio. As the calculations involve binomial coefficients and \eqref{eq::revenue} involves fractions, it is not straightforward to find $L^*$. Instead, here, we provide a much simpler method to obtain the optimal $L$. To do so, we resort to the linearization method of the OSM MDP which is presented in Appendix~\ref{sec::app-existing}, and use it in a narrow-sense. Note that, every time a block is mined, the adversary faces a decision as to whether it should release $A$-chain and end the cycle, or continue. Clearly, if $A-H>1$, the decision should be to continue the attack cycle as the adversary can gain additional advantage at no-cost in these situations. On the other hand, if $A=H+1$, the risk is imminent as the adversary loses the advantage of overriding the $H$-chain if the next arrival is honest. As a result, we can narrow down the decision problem to a specific case where $A=H+1$ and analyze the situation therein via the linearization method. 

To that end, we define $u$ and $v_a$ as
\begin{align}
    u=(1-\gamma), \qquad 
    v_{a}=\frac{1}{u}\left(1-\gamma\frac{1-\rho_{a}}{1-2\alpha}\right).
\end{align}

Next, using the definitions above and intermediary results in Appendix~\ref{sec::app-l-stub}, we provide the following quasiconcavity result related to the $L$-stubborn mining.

\begin{theorem}\label{thm::quasiconcavity} \textbf{(Quasiconcavity)}
    $\rho_{L}\geq\rho_{L+1}$ implies $\rho_{a}\geq\rho_{a+1}$ for any $a\geq L$. Similarly, $\rho_{M}\geq\rho_{M-1}$ implies $\rho_{a}\geq\rho_{a-1}$ for any $a\leq M$.    
\end{theorem}

Theorem~\ref{thm::quasiconcavity} essentially says that $\rho_L$ is quasiconcave for integer $L$. We relegate the proof of Theorem~\ref{thm::quasiconcavity} to the Appendix~\ref{sec::app-l-stub} as it requires some intermediary definitions and results. These results also shed light on a simple algorithm that can be used to find $L^*$.

Next, we provide Algorithm~\ref{alg::cap} that finds the optimal $L^*$ of $L$-stubborn mining. For that, let $\rho_1$, $\rho_2$ and $\rho_\infty$ be defined as in Definition~\ref{def::rev_ratios}, which are the revenue ratios of honest mining, selfish mining \cite{selfish-mining}, and equal-fork stubborn mining \cite{stubborn-mining,catalan-stubborn-grunspan}, respectively. Note that each of these values can be obtained by \eqref{eq::revenue}, where for $\rho_{\infty}$, $L_s$ can be set to zero as the $H$-chain will catch the $A$-chain eventually, \textit{i.e.}, no attack ends successfully, but the revenue comes from matching. In Appendix~\ref{sec::app-l-stub}, we use the intermediary results to show that Algorithm~\ref{alg::cap} outputs $L^*$, which essentially solves the decision problem mentioned earlier.

\begin{algorithm}[t!]
    \caption{Optimal $L^*$ algorithm}\label{alg::cap}
    \begin{algorithmic}[1]
    \STATE $i\gets 1$
    \STATE $L^{(1)}\gets \arg\max(\rho_1,\rho_2,\rho_{\infty})$
    \IF{$v_{L^{(1)}}\leq 0$}
        \RETURN $\infty$
    \ENDIF
    \WHILE{$L^{(1)}>1$}
        \STATE $L^{(i+1)}\gets \ceil{\frac{\log v_{L^{(i)}}}{\log u}}$
    \IF{$L^{(i+1)} = L^{(i)}$}
        \RETURN $L^{(i)}$
    \ELSE
        \STATE $i \gets i + 1$
    \ENDIF
    \ENDWHILE
    \RETURN $L^{(i)}$
    \end{algorithmic}
\end{algorithm}

\subsection{$(k+1)$-Stubborn Mining and Double-Spending}
In the analysis of double-spending, we focus on $L=(k+1)$-stubborn mining strategy for the ease of the analysis. However, we note that, the choice of $L$ does not need to equal $k+1$ for the adversary to get free shots at double-spending. For example, the attacker can try to use the optimal value $L=L^*$ that maximizes the revenue ratio of coinbase rewards and still some attack cycles result in double-spending. More specifically, if
\begin{enumerate}
    \item $L< k+1$, successful attack cycles that reach $A=k+1$ may still result in double-spending. 
    \item $L> k+1$, both successful and unsuccessful attack cycles may result in double-spending. 
\end{enumerate}
We do not focus on the double-spending analysis of these two cases to avoid a complicated analysis. We pick $L=k+1$ when focusing on the double-spending since under the $(k+1)$-stubborn strategy the aim of the attacker is to undo an $H$ chain with length $H\geq k$ whenever possible. As a result, $(k+1)$-stubborn mining can be seen as a sustained double-spending attack on the first block height in each attack cycle until it succeeds in double-spending. For such an attack, we already provided the formula to calculate the revenue ratio of the $(k+1)$-stubborn mining. Here, we provide the probability that an attack cycle results in \textbf{Double-Spending} event.

Note that, even when $L=k+1$, a successful attack cycle does not immediately imply a successful double-spending. It only implies that the $A$-chain will make it into the offset chain eventually. Similarly, an unsuccessful attack cycle does not immediately imply that the first block of the $H$-chain, $\mathsf{B}$, will be confirmed. Next, we provide the probability of each event regarding $\mathsf{B}$ and $\mathsf{B'}$ for $(k+1)$-stubborn mining with $k$-confirmation rule.

\begin{theorem}\label{thm::db_sp} \textbf{(Double Spend)}
    Under $k$-confirmation rule, each attack cycle of $(k+1)$-stubborn mining ends with Service, Move-Funds (MF) or Double-Spending (DS) on the first block height with probabilities,
    \begin{align}
        P(\textbf{DS})=&\sum_{m=0}^{k}P_s(k+1,m)(1-\gamma)^{m-\mathbbm{1}_{m=k}}\nonumber\\&+P_{u}(k)(1-\gamma)^{k-1}\gamma, \label{eq::P_db_sp} \\
        P(\textbf{Service})&=\sum_{m=0}^{k}P_u(m)(1-\gamma)^m, \\
        P(\textbf{MF})&=1-P(\textbf{DS})-P(\textbf{Service}).
    \end{align}
\end{theorem}

\begin{Proof}
    For Double-Spending to happen, one possibility is, we will observe $A>k$ and $H\leq k$ at some point during the cycle. The first time this happens during the cycle is the first time $A=k+1$ and $H=m$ which happens w.p. $P_s(k+1,m)$. Until this point, the adversary matches every one of the $m$ honest blocks. From this point onwards, the adversary does not need to match the honest blocks as the strategy will eventually release the longer $A$-chain. Hence, $\mathsf{B}$ stays in the $H$-chain if none of the $m$ matching adversarial blocks make it into the prefix of the honest block at height $H$, which happens w.p. $(1-\gamma)^{m-\mathbbm{1}_{m=k}}$. Note that, if $H=k$, then it does not matter whether prefix chain is after the $k$th block or not, hence $\mathbbm{1}_{m=k}$ at the exponent. Another possibility for Double-Spending is, if $H$-chain catches $A$-chain with both having length $k$ and the next block is honest which has probability $P_{u}(k)$. In this situation, if $H$-chain switches its prefix on the last honest block, which happens w.p. $(1-\gamma)^{k-1}\gamma$, then $\mathsf{B}$ gets discarded despite being confirmed earlier.   For Service to happen, we will observe $L \geq H=A+1=m+1$, which ends the cycle with $H$-chain making it into the offset chain which happens w.p. $P_{u}(m)$. Further, $\mathsf{B}$ has to be in the prefix of the honest block at height $m+1$, which explains the term $(1-\gamma)^{m}$.
\end{Proof}

Here, we remind the reader that $\rho_{k+1}$ is the revenue ratio of the adversary when it employs the $(k+1)$-stubborn mining strategy and $\Bar{L}=\sup\{L:\rho_L\geq\alpha\}$, \textit{i.e.}, the largest value of $L$ such that the revenue ratio of $L$-stubborn mining is higher than the honest mining strategy. Hence, due to the quasiconcavity proven in Theorem~\ref{thm::quasiconcavity}, under any $k$-confirmation rule with $k<\Bar{L}$ is compromised by $(k+1)$-stubborn mining strategy since the adversary gains more revenue ratio by employing this strategy compared to the revenue ratio of the honest mining strategy, \textit{i.e.}, $\rho_{k+1}\geq\alpha$, and at the same time it can double-spend at no-cost. In other words, for a given $\alpha$ and $\gamma$, $k$-confirmation rule should be at least $k\geq \Bar{L}$ to make sure that double-spending attack has loss of revenue. Note that, if $\rho_{\infty}\geq\alpha$, no $k$-confirmation is safe as double-spending has zero-cost. From this perspective, the safety of a blockchain system fails well before adversarial proportion reaches $\alpha$. We investigate this phenomenon further in the numerical results section by providing $L^*$ and $\Bar{L}$ for each $\alpha$ and $\gamma$.

\subsection{Combined Revenue Ratio}\label{sec::comb_revenue}
In this subsection, we provide a formula that incorporates the profit of double-spending into the revenue ratio the adversary makes from $(k+1)$-stubborn mining. To do so, in each attack cycle, we assume that the profit of successful double-spend is equal to $\mathcal{R}$ units of block rewards. More specifically, replacement of a confirmed target block $\mathsf{B}$ on the first block height of the attack cycle with another block $\mathsf{B'}$ is considered as a double-spend event and brings additional $\mathcal{R}$ units of block rewards. Other considerations are left as simple extensions which the reader can easily calculate given the approaches we provide here.

\begin{corollary}\label{thm::comb_revenue} \textbf{(Combined Revenue Ratio)}
    The combined revenue ratio of $(k+1)$-stubborn mining and the associated double-spending is given as
    \begin{align}
        \rho'_{(k+1)}(\mathcal{R})=\frac{(k+1)_s+(k+1)_{u,a}+\mathcal{R}\cdot P(\textbf{DS})}{(k+1)_s+(k+1)_u}, \label{eq::comb_revenue}
    \end{align}
    where $(k+1)_s$, $(k+1)_{u,a}$ and $(k+1)_u$ are defined as in Theorem~\ref{thm::revenue}, where we pick $L=(k+1)$ and P(\textbf{DS}) is defined as in Theorem~\ref{thm::db_sp}.
\end{corollary}

We define 
\begin{align}
    R^*_{k}&=\inf\{\mathcal{R}:\rho'_{k+1}(\mathcal{R})\geq\alpha\},
\end{align}
which, essentially, gives the minimum double-spending reward value required for the scheme to be profitable on average. We provide $R^*_{k}$ for each $\alpha$ and $\gamma$ value for Bitcoin, where $k=6$, in numerical results later.

Next, we provide another approach that incorporates the profit coming from double-spending when considering the $(k+1)$-stubborn mining attack provided here. As mentioned in \cite{rosenfeld2014analysis}, the value of service provided might be below the fee paid. Hence, we assume that the value of service provided by the merchant equals $V$ and the adversary pays $F$ fee for the service with $F\geq V$, where both values are in block reward units. In this case, on average, the $(k+1)$-stubborn mining attack is more profitable than honest mining if
\begin{align}
    &(k+1)_s+(k+1)_{u,a}+V\cdot P(\textbf{Double-Spending}) \nonumber\\ & \geq \alpha\cdot \left((k+1)_s+(k+1)_{u}\right)+(F-V)\cdot P(\textbf{Service}).
\end{align}
The reasoning for this is as follows: Out of every $\left((k+1)_s+(k+1)_{u}\right)$ blocks that make it into the prefix chain, if the adversary followed the honest protocol, $\alpha$ fraction would be its revenue ratio. However, with the $(k+1)$-stubborn mining protocol, $(k+1)_s+(k+1)_{u,a}$ blocks make it into the prefix chain. Further, the double-spending results in $V$ extra rewards for the adversary. However, if only \textbf{Service} happens, $F-V$ amount of rewards is lost.

\begin{algorithm}[t!]
    \caption{Optimal $S^*$ algorithm}\label{alg::cap_stealth}
    \begin{algorithmic}[1]
    \STATE $i\gets 1$
    \STATE $S^{(1)}\gets \arg\max(\sigma_1,\sigma_2)$
    \WHILE{$S^{(1)}>1$}
        \STATE $S^{(i+1)}\gets \ceil{f(\sigma_{S(i)})}$
    \IF{$S^{(i+1)} = S^{(i)}$}
        \RETURN $S^{(i)}$
    \ELSE
        \STATE $i \gets i + 1$
    \ENDIF
    \ENDWHILE
    \RETURN $S^{(i)}$
    \end{algorithmic}
\end{algorithm}

\section{Analysis of $S$-Stealth Mining}\label{sec::stealth}
An interesting observation of the results in Theorem~\ref{thm::db_sp} is the fact that $P(\textbf{Double-Spending})$ is decreasing in $\gamma$ due to the fact that, even if the attack cycles end successfully, $\gamma$ increases the probability that the prefix of $H$-chain switches to $\mathsf{B'}$ before $\mathsf{B}$ is confirmed, because the adversary matches each honest block before $A=L$. Thus, for a given $\alpha$, while $\gamma$ increases, both $P(\textbf{Double-Spending})$ and $P(\textbf{Service})$ decrease and $P(\textbf{Move-Funds})$ increases. Further, even if the $H$-chain still contains $\mathsf{B}$ in its prefix, every time an honest block is mined before $A\geq L$, the adversary matches it, \textit{i.e.}, releases a competing block on the same height which contains $\mathsf{B'}$ in its prefix, which would make the merchant providing services related to $\mathsf{B}$ suspicious. In this section, we provide a modified version of $L$-stubborn mining, which we call $S$-stealth mining, aiming to make the detection of the double-spending attack difficult and to increase $P(\textbf{Double-Spending})$ in each attack cycle.

In $S$-stealth mining, the strategy of keeping track of the length of $A$-chain is still the same, however, matching only happens if $A=H=S-1$. In other words, whenever the honest miners mine a new block on a height $H<S-1$, the adversary does not match the honest block. On the other hand, if $A=S-1$ and $H=S-2$ and the next block is honest on height $S-1$, the adversary matches this block.

\begin{theorem}\label{thm::revenue_stealth} \textbf{(Revenue Ratio)}
    The revenue ratio of $S$-stealth mining is given as
    \begin{align}
        \sigma_S=\frac{S_s+S_{u,s}}{S_s+S_u}, \label{eq::stealth_revenue}
    \end{align}
    where
    \begin{align}
        S_s&=\sum_{m=0}^{S-1} P_s(S,m)\left(S+\frac{(S-m-1)\alpha}{1-2\alpha}\right),\\
        S_{u,s}&= P_u(S-1) \gamma  (S-1),\\
        S_{u}&=\sum_{n=0}^{S-1}P_u(n) (n+1). 
    \end{align}
\end{theorem}

\begin{Proof}
For a successful attack, we follow the same arguments provided in the proof of Theorem~\ref{thm::revenue}. When an attack is unsuccessful, adversarial blocks make it into the offset chain only if the $H$-chain has reached length $S-1$, and the next honest block on height $S$ switches its prefix (w.p. $\gamma$) to the $A$-chain which has length $S-1$.
\end{Proof}

We note that $\sigma_1=\rho_1=\alpha$ and $\sigma_2=\rho_2$, \textit{i.e.}, $2$-stealth mining is the same algorithm as selfish mining of \cite{selfish-mining}. Further, $\sigma_{\infty}=0$ as there is no matching when $S=\infty$ and eventually $H$-chain will overtake in every cycle.

\subsection{Optimizing the Revenue Ratio}
Here, we first provide the theorem stating the quasiconcavity of $S$-stealth mining. Even though the goal in $S$-stealth mining is to increase the double-spending chance instead of focusing on the revenue ratio, for the sake of completeness, for a given $(\alpha,\gamma)$ pair, we also provide an algorithm to find $S^*$, \textit{i.e.}, the integer value of $S$ maximizing the revenue ratio. The proofs are provided in Appendix~\ref{sec::app-s-steal}.

\begin{theorem}\label{thm::quasiconcavity_stealth} \textbf{(Quasiconcavity)}
    $\sigma_{S}\geq\sigma_{S+1}$ implies $\sigma_{a}\geq\sigma_{a+1}$ for any $a\geq S$. Similarly, $\sigma_{R}\geq\sigma_{R-1}$ implies $\sigma_{a}\geq\sigma_{a-1}$ for any $a\leq R$.
\end{theorem}

Next, we provide Algorithm~\ref{alg::cap_stealth} below, which returns $S^{*}$, where
\begin{align}
    f(\sigma)=&\sup\bigg\{x\in\mathbb{R}:1-\frac{\gamma(1-2\alpha)}{\alpha(1-\alpha)}\frac{x^2-1}{2(2x-1)}\nonumber\\&-(1-2\alpha)(x+1-\gamma x)\geq \sigma\bigg\},
\end{align}
and note that $\sigma_1=\alpha$ and $\sigma_2=\rho_2$, which is provided in \eqref{eq::self_formula}.

\begin{table}[t]    
\begin{center}
\begin{tabular}{||c c c c c||} 
 \hline
 $\alpha$ & $\rho_2$\cite{selfish-mining} & $\rho_{L^*}$  & $\rho_{OSM}$\cite{optimal-selfish} & $\rho_{PTO}$\cite{prob-selfish-mdp-method} \\ [0.5ex] 
 \hline
 1/3 & 0.33269 & 0.33333 & 0.33705 & 0.33705 \\ 
 \hline
 0.35 & 0.36651 & 0.36651 & 0.37077 & 0.37077 \\
 \hline
 0.375 & 0.42118 & 0.42118 & 0.42600 & 0.42600 \\
 \hline
 0.4 & 0.48372 & 0.48372 & 0.48866 & 0.48866 \\
 \hline
 0.425 & 0.55801 & 0.55801  & 0.56808 & 0.56809 \\
 \hline
 0.45 & 0.65177& 0.66248  & 0.66891 & 0.66894 \\
 \hline
 0.475 & 0.78255 & 0.80043  & 0.80172 & 0.80184\\  
 \hline
\end{tabular}
\end{center}
\caption{ Comparison between the revenue ratio of $L^*$-stubborn mining and MDP models \cite{prob-selfish-mdp-method, optimal-selfish} for $\gamma=0$.}
\label{tab::compare_with_mdp}
\end{table}

\subsection{$(k+1)$-Stealth Mining and Double-Spending}

Similar to the $(k+1)$-stubborn mining, $(k+1)$-stealth mining can also be seen as a sustained double-spending attack until it succeeds in double-spending. The difference is that the attacker here gives up some coinbase revenue by avoiding the \textit{matching} actions, which allows the attack to be not exposed as well as increases the double-spending probability. Notice that $(k+1)$-stealth mining is essentially the private attack of Nakamoto \cite{btc-whitepaper} with the exception that, here, the adversary gives up pursuing the attack if it falls behind before reaching length $k$ to avoid revenue losses. If it reaches length $k$ but the honest chain catches it at length $k$, the network split $\gamma$ could still help the adversary to double-spend. However, as matching only happens at length $k$, there is no Move-Funds event possible in $(k+1)$-stealth mining. We already provided the formula to calculate the revenue ratio of the $(k+1)$-stealth mining. Here, we provide the probability that an attack cycle results in \textbf{Double-Spending} event.

\begin{table*}[t!] 

\begin{center}\footnotesize
\begin{tabular}{|l || l | l || l | l || l | l || l | l || l | l|} 
 \hline
  & \multicolumn{2}{c||}{ $\gamma=0.2$}& \multicolumn{2}{c||}{ $\gamma=0.4$}& \multicolumn{2}{c||}{ $\gamma=0.5$}& \multicolumn{2}{c||}{ $\gamma=0.6$}& \multicolumn{2}{c|}{ $\gamma=0.8$}   
 \\\hline
 $\alpha$  & $\rho_{L^*}$& $\rho_{OSM}$ & $\rho_{L^*}$& $\rho_{OSM}$ & $\rho_{L^*}$& $\rho_{OSM}$ & $\rho_{L^*}$& $\rho_{OSM}$ & $\rho_{L^*}$& $\rho_{OSM}$   \\ [0.5ex]
 \hline
 0.1 & 0.1 & 0.1 & 0.1 & 0.1 & 0.1 & 0.1 & 0.1 & 0.1 & 0.1 & 0.1\\ 
 \hline
 0.15 & 0.15 & 0.15 & 0.15 & 0.15 & 0.15 & 0.15 & 0.15 &0.15 & 0.151 & 0.151\\
 \hline
 0.2 & 0.2 & 0.2 & 0.2 & 0.2 & 0.2 & 0.2 & 0.2 & 0.2 & 0.218 & 0.218\\
 \hline
 0.25 & 0.25 & 0.25 & 0.25 & 0.25 & 0.25 & 0.25 & 0.261 & 0.261 & 0.297 & 0.297\\
 \hline
 0.3 & 0.3 & 0.3  & 0.316 & 0.316 & 0.327 & 0.327 & 0.342 & 0.344 & 0.390 & 0.390 \\
 \hline
 0.35 & 0.386 & 0.388 & 0.407 & 0.410 & 0.425 & 0.430 & 0.453 & 0.456 & 0.502 & 0.502 \\
 \hline
 0.4 & 0.502 & 0.511 & 0.536 & 0.547& 0.566 & 0.573 & 0.594 & 0.597 & 0.636 & 0.637\\  
 \hline
 0.45 & 0.686 & 0.700  & 0.729 & 0.740& 0.753 & 0.760 & 0.772 & 0.776 & 0.799 & 0.800\\  
 \hline
\end{tabular}
\end{center}
\caption{ Comparison between the revenue ratio of $L^*$-stubborn mining and OSM MDP for varying $\gamma$.}
\label{tab::compare_with_mdp_vary_gamma}
\end{table*}

\begin{theorem}\label{thm::stealth_db_sp} \textbf{(Double Spend)}
    Under the $k$-confirmation rule, each attack cycle of $(k+1)$-stealth mining ends with Service or Double-Spending (DS) on the first block height with probabilities
    \begin{align}
        P(\textbf{DS})&=\sum_{m=0}^{k}P_s(k+1,m)+P_{u}(k)\gamma, \label{eq::stealth_P_db_sp}\\
        P(\textbf{Service})&=1-P(\textbf{DS}).
    \end{align}
\end{theorem}

\begin{Proof}
    Contrary to $(k+1)$-stubborn mining, a successful cycle of $(k+1)$-stealth mining always results in double-spending since the $H$-chain always keeps $\mathsf{B}$ in its prefix and the $A$-chain undoes it eventually. On the other hand, if an attack ends unsuccessfully with $H=(k+1)$, the honest chain might switch from $\mathsf{B}$ to $\mathsf{B'}$ right at the end when the honest block at height $k+1$ is mined. This would imply that at first $\mathsf{B}$ is confirmed when the $H$-chain reaches length $k$ and the next honest block switches prefix and $\mathsf{B'}$ is confirmed, resulting in a double-spend. 
\end{Proof}

The formula we provide in Theorem~\ref{thm::revenue_stealth} can be used to determine the revenue ratio  of $(k+1)$-stealth mining, \textit{i.e.}, $\sigma_{k+1}$, which is no different than the private attack with no-trail. If $\sigma_{k+1}\geq \alpha$, this would imply that trying private attack with no-trail comes at no-cost, \textit{i.e.}, the revenue ratio is still above or equal to $\alpha$. Hence, similar to the case of $L$-stubborn mining, we remind the reader that $\Bar{S}=\sup\{S:\sigma_S\geq\alpha\}$, \textit{i.e.}, the largest value of $S$ such that the revenue ratio of $S$-stealth mining is higher than the honest mining strategy. Hence, due to the quasiconcavity proven in Theorem~\ref{thm::quasiconcavity_stealth}, under any $k$-confirmation rule with $k<\Bar{S}$ is compromised by $(k+1)$-stealth mining strategy since the adversary gains more revenue ratio by employing this strategy compared to the revenue ratio of the honest mining strategy, \textit{i.e.}, $\sigma_{k+1}\geq\alpha$, and at the same time it can double-spend at no-cost. We provide the associated values of $S^*$ and $\Bar{S}$ for given $\alpha$ and $\gamma$ in the numerical results section.

\subsection{Combined Revenue Ratio}\label{sec::comb_revenue_stealth}
Here, we provide the approach where each successful double-spend event on the target block of an attack cycle results in $\mathcal{R}$ additional block rewards for the adversary. In this case, the revenue ratio is provided in Corollary~\ref{thm::comb_revenue_stealth}. The metric introduced in Corollary~\ref{thm::comb_revenue_stealth} is similar to the \textit{subversion gain} defined as in \cite{preneel_common_metrics}.

\begin{corollary}\label{thm::comb_revenue_stealth} \textbf{(Combined Revenue Ratio)}
    The combined revenue ratio of $(k+1)$-stealth mining and the associated double-spending is given as
    \begin{align}
        \sigma'_{(k+1)}(\mathcal{R})=\frac{(k+1)_s+(k+1)_{u,s}+\mathcal{R}\cdot P(\textbf{DS})}{(k+1)_s+(k+1)_u}, \label{eq::comb_revenue_stealth}
    \end{align}
    where $(k+1)_s$ and $(k+1)_u$ are defined as in Theorem~\ref{thm::revenue_stealth}, where we pick $S=(k+1)$ and P(\textbf{DS}) is defined as in Theorem~\ref{thm::stealth_db_sp}.
\end{corollary}

We define the minimum double-spend reward value needed for a profitable scheme as
\begin{align}
    R^*_{k}&=\inf\{\mathcal{R}:\sigma'_{k+1}(\mathcal{R})\geq\alpha\}. \label{eq::subversion_bounty}
\end{align}
Note that, $R^*_{k}$, defined here, is closely related to the \textit{subversion bounty} defined in \cite[Section IV.B.2]{preneel_common_metrics}. As matched honest blocks are not considered for double-spending since they can make the merchant suspicious, \textit{subversion gain} and \textit{subversion bounty} of stealth mining is much closer to the definition of \cite{preneel_common_metrics} than the corresponding results provided for stubborn mining in Section~\ref{sec::comb_revenue}.

If we assume that each successful cycle results in only a single double-spending event with a service value $V$ and payment $F$ with $F\geq V$ both in block rewards units, then, on average, the $(k+1)$-stealth mining is more profitable than honest mining if 
\begin{align}
    &(k+1)_s+(k+1)_{u,s}+V \nonumber\\&\geq \alpha\cdot \left((k+1)_s+(k+1)_{u}\right)+F\cdot P(\textbf{Service}),
\end{align}
where all values of $(k+1)_s$, $(k+1)_{u,s}$, $(k+1)_{u}$, $P(\textbf{Service})$ are provided as in Theorem~\ref{thm::revenue_stealth} and Theorem~\ref{thm::stealth_db_sp}.

\begin{figure*}[ht!]
\captionsetup[subfigure]{aboveskip=1pt,belowskip=9pt}
     \centering
     \begin{subfigure}[b]{0.31\textwidth}
         \centering
         \includegraphics[width=\textwidth]{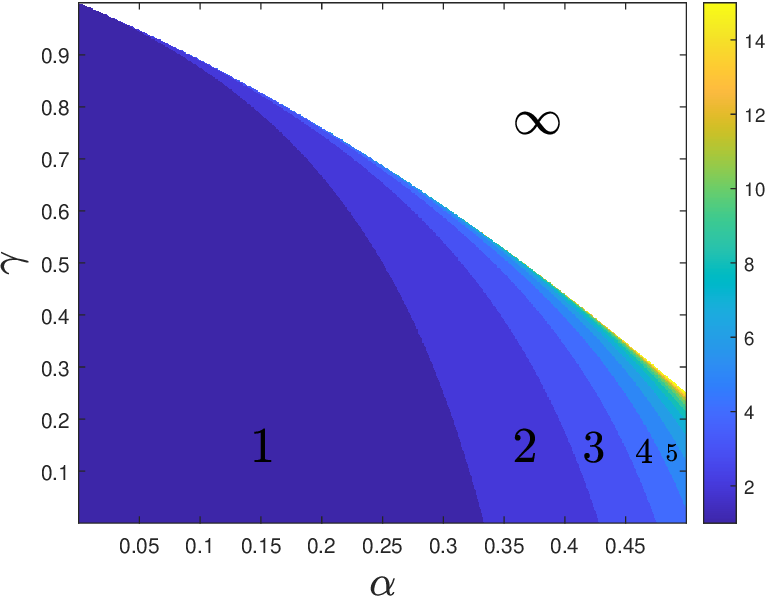}
         \caption{$L^*$}
         \label{fig::best_L}
     \end{subfigure}
    \hfill
    \begin{subfigure}[b]{0.31\textwidth}
         \centering
         \includegraphics[width=\textwidth]{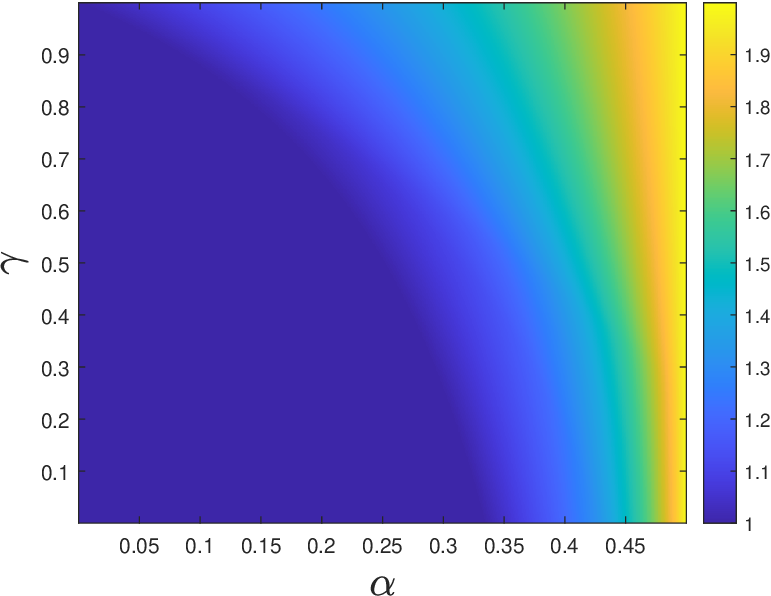}
         \caption{$\frac{\rho_{L^*}}{\alpha}$}
         \label{fig::L_revenue_prop}
    \end{subfigure}
    \hfill
    \begin{subfigure}[b]{0.31\textwidth}
         \centering
         \includegraphics[width=\textwidth]{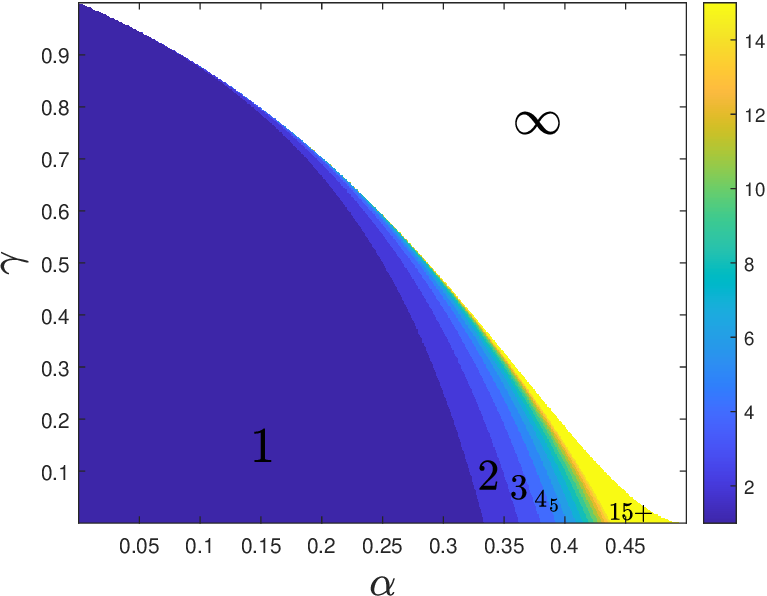}
         \caption{$\Bar{L}$}
         \label{fig::Bar_L}
    \end{subfigure}
    \hfill
    \begin{subfigure}[b]{0.31\textwidth}
         \centering
         \includegraphics[width=\textwidth]{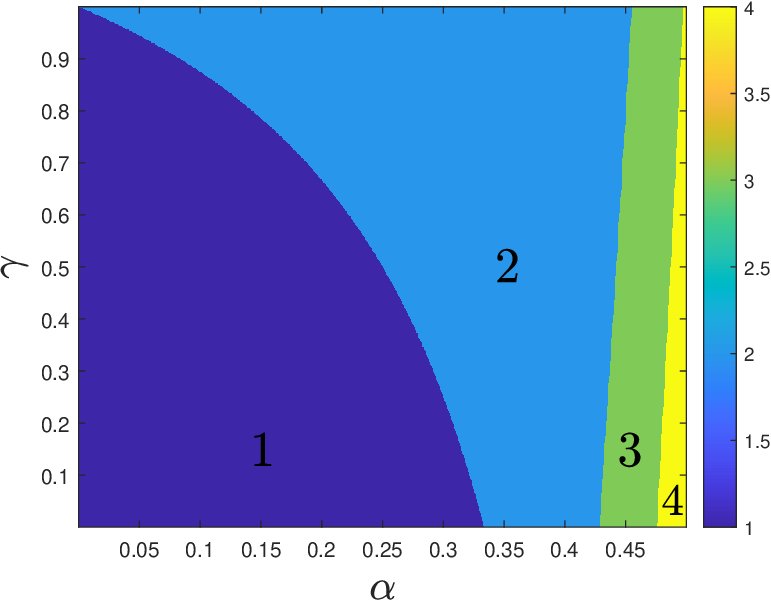}
         \caption{$S^*$}
         \label{fig::best_L_stealth}
    \end{subfigure}
    \hfill
    \begin{subfigure}[b]{0.31\textwidth}
         \centering
         \includegraphics[width=\textwidth]{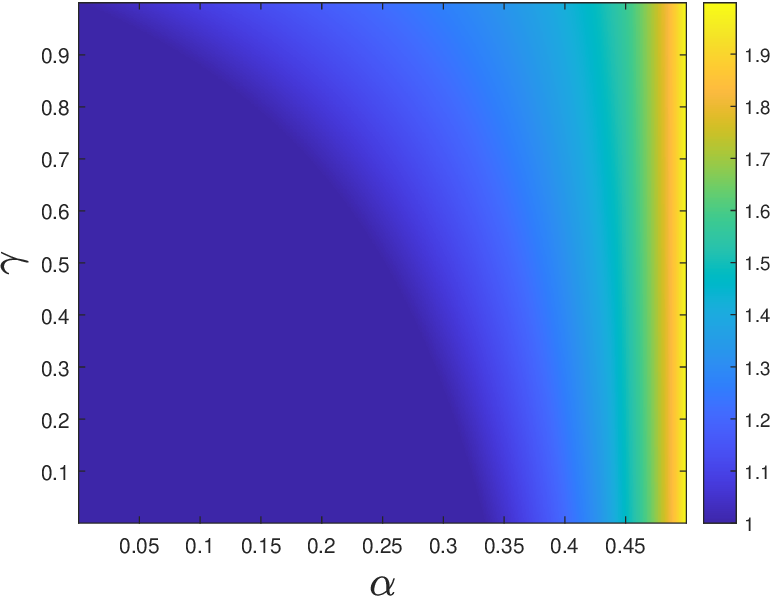}
         \caption{$\frac{\sigma_{S^*}}{\alpha}$}
        \label{fig::L_revenue_prop_stealth}
    \end{subfigure}
    \hfill
    \begin{subfigure}[b]{0.31\textwidth}
         \centering
         \includegraphics[width=\textwidth]{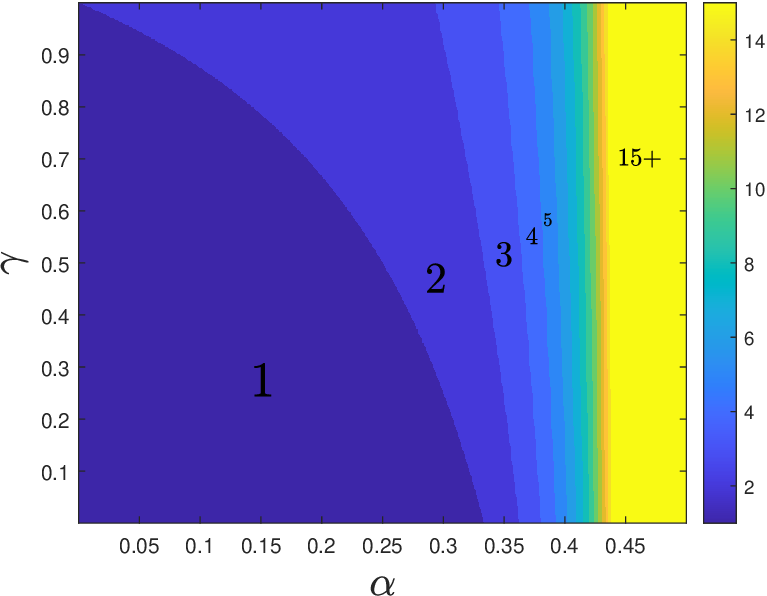}
         \caption{$\Bar{S}$}
         \label{fig::Bar_L_stealth}
    \end{subfigure}
     
    \caption{$L$-stubborn mining and $S$-stealth mining.}
	\label{fig::regions}
\end{figure*}

\section{Numerical Results}\label{sec::numerical}
We start by evaluating $L^*$ for each value of $\alpha$ and $\gamma$ using Algorithm~\ref{alg::cap}. The results are provided in \figref{fig::best_L}. Note that the boundary between $L^{*}=1$ and $L^{*}=2$, \textit{i.e.}, the honest mining and selfish mining, was already provided in \cite{selfish-mining}. Our results here show new boundaries by providing regions where waiting for the private branch to extend further is beneficial in terms of the revenue ratio. Next, in \figref{fig::L_revenue_prop} for each $\alpha$ and $\gamma$, we provide the normalized revenue ratio of optimal $L^{*}$-stubborn mining attack compared to honest mining. For all values of $\gamma$ the normalized revenue ratio starts as $1$ from $\alpha=0$ and reaches $2$ at $\alpha=0.5$, as expected since at $\alpha=0.5$, the adversary virtually creates all the blocks in the offset chain. However, as $\gamma$ increases, the ratio increases faster initially. 

In Table \ref{tab::compare_with_mdp}, we compare the optimal revenue ratio obtained from $L^*$-stubborn mining with the revenue ratio obtained from selfish mining \cite{selfish-mining} (where $L=2$) and OSM MDP \cite{optimal-selfish} and Probabilistic Termination Optimization (PTO) MDP \cite{prob-selfish-mdp-method} to demonstrate that our model performs close to the $\epsilon$-optimal selfish mining strategies. We use the data from \cite[Table 1]{prob-selfish-mdp-method}, where $\gamma=0$ and both MDP models are limited to maximum fork length of $95$. In Table \ref{tab::compare_with_mdp_vary_gamma}, we compare $\rho_{L^*}$ with $\rho_{OSM}$ of \cite{optimal-selfish} for varying $\alpha$ and $\gamma\neq0$ values, where we use the implementation of \cite{mitsuhamizu} for the OSM MDP with a maximum fork length of $80$. It is quite clear that our simplified $L^*$-stubborn strategy results in reward ratio $\rho_{L^{*}}$, that is usually within $10^{-2}$ of the reward ratio $\rho_{OSM}$ for varying values of $\gamma$. As we mentioned earlier, the small difference can be attributed to the fact that, in MDP models, when the adversarial chain falls behind honest chain while having a length close to but smaller than $L$, the adversary does not immediately adopt the honest chain but instead tries to catch up to the honest chain which brings some additional revenue, albeit rather small.

Finally, in \figref{fig::Bar_L}, we provide $\Bar{L}$ for each $\alpha$ and $\gamma$, which we believe has a significant meaning related to $(k+1)$-confirmation rule. As we proved that $\rho_L$ has a quasiconcave behavior, if $\Bar{L}>k$, then $(k+1)$-stubborn mining is more profitable than honest mining, hence following the $(k+1)$-stubborn mining allows double-spending at no-cost. This in turn means, for each $\alpha$ and $\gamma$, $k$-confirmation rule has to be set according to $k\geq \Bar{L}(\alpha,\gamma)$ to minimize the risk of double-spending that comes at no-cost to the attacker.
\begin{figure*}[ht!]
     \centering
     \begin{subfigure}[b]{0.24\textwidth}
         \centering
         \includegraphics[width=\textwidth]{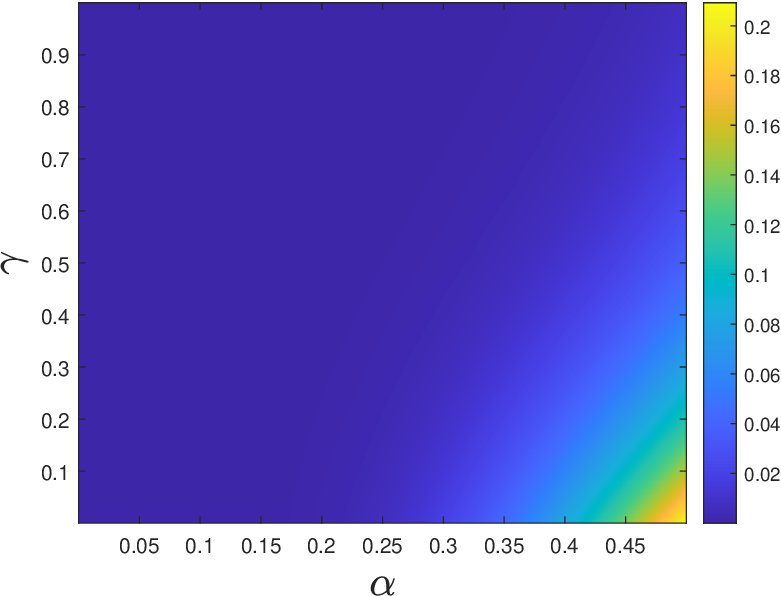}
         \caption{P(\textbf{Double-Spending})}
         \label{fig::btc_double_sp}
     \end{subfigure}
     \hfill
     \begin{subfigure}[b]{0.24\textwidth}
         \centering
         \includegraphics[width=\textwidth]{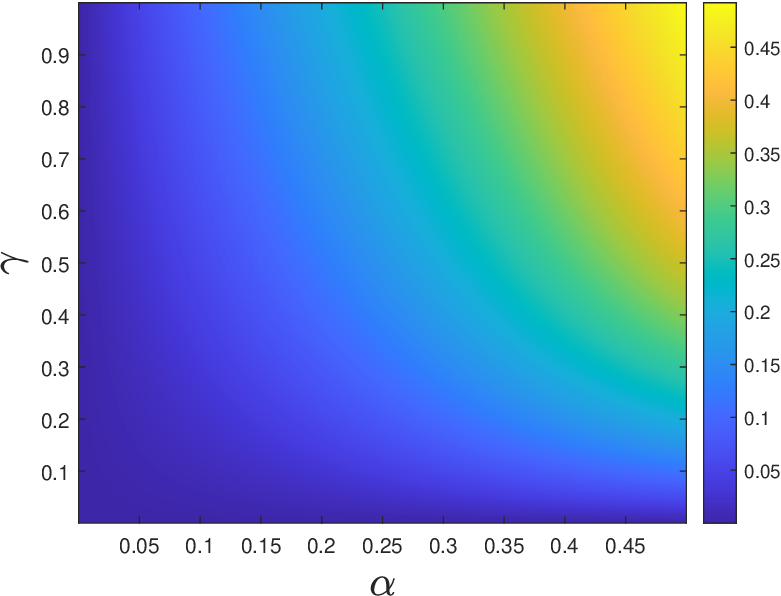}
         \caption{P(\textbf{Move-Funds})}
         \label{fig::btc_move}
     \end{subfigure}
     \hfill
     \begin{subfigure}[b]{0.24\textwidth}
         \centering
         \includegraphics[width=\textwidth]{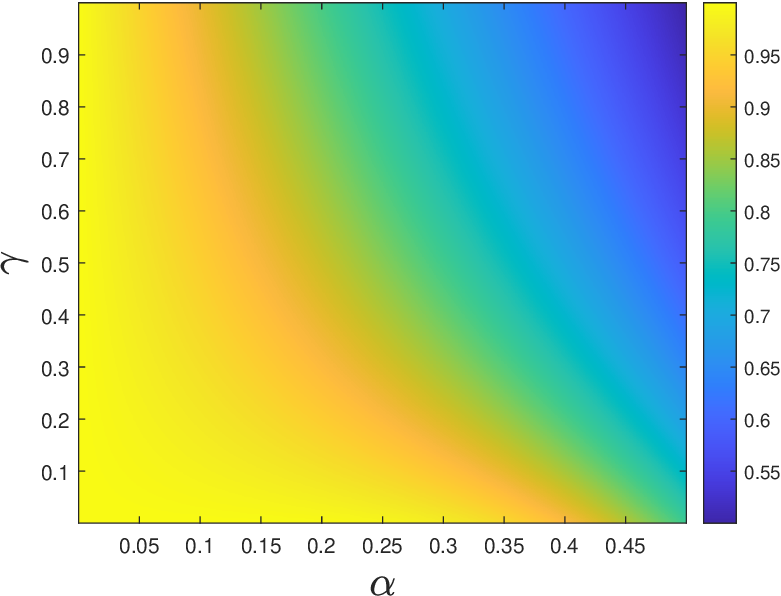}
         \caption{P(\textbf{Service})}
         \label{fig::btc_service}
     \end{subfigure}
     \hfill
     \begin{subfigure}[b]{0.235\textwidth}
         \centering
         \includegraphics[width=\textwidth]{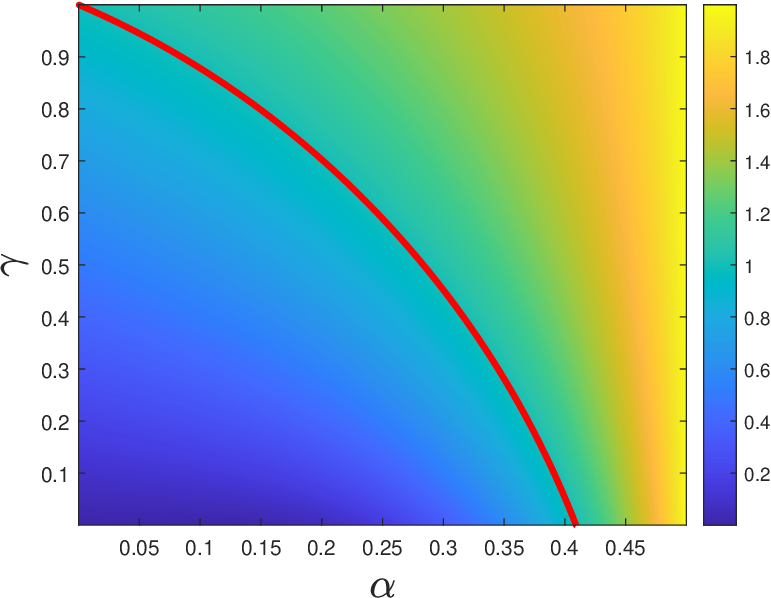}
         \caption{$\frac{\rho_{k+1}}{\alpha}$}
         \label{fig::btc_revenue_scaled}
     \end{subfigure}
     
    \caption{$(k+1)$-stubborn mining with $k=6$.}
	\label{fig::k_plus_1_probs}
\end{figure*}
Similar to the results of $L$-stubborn mining, \figref{fig::regions} also contains related results for $S$-stealth mining. Observing the $S^{*}$ provided in \figref{fig::best_L_stealth}, it is clear that $S^{*}\leq L^{*}$ which is straightforward from the analysis provided earlier, but also $S^{*}\leq 4$ for $\alpha<0.5$ and $\gamma<1$. In other words, in stealth mining, the adversary cannot act in a stubborn manner forever, whereas acting stubborn no matter what the length of the adversarial branch is the best strategy to maximize the revenue ratio for $L$-stubborn mining in the white region of \figref{fig::best_L}. In \figref{fig::L_revenue_prop_stealth}, we provide the normalized revenue ratio of optimal $S^{*}$-stealth mining attack. Although it starts as $1$ from $\alpha=0$ and reaches $2$ at $\alpha=0.5$ similar to the $L^{*}$-stubborn mining, the rise in the revenue ratio is not as steep initially. Further, as expected, the revenue difference between $L^{*}$-stubborn mining and $S^{*}$-stealth mining increases with $\gamma$, as matching is only present in the last step in stealth mining. Finally, \figref{fig::Bar_L_stealth} shows the values of $\Bar{S}$, where we have $\Bar{S}<\infty$ for all values of $\alpha<0.5$ and $\gamma<1$, since $\sigma_{\infty}=0$.

In \figref{fig::k_plus_1_probs} we provide the probabilities of the events \textbf{Double-Spending}, \textbf{Move-Funds} and \textbf{Service} for each attack cycle of $(k+1)$-stubborn mining where we pick $k=6$ as in Bitcoin. We also provide normalized revenue ratio of $(k+1)$-stubborn mining in \figref{fig::btc_revenue_scaled} where the red curve highlights the boundary where the ratio rises above 1. It is clear that the red curve is the same as the boundary between $\Bar{L}=6$ and $\Bar{L}=7$ provided in \figref{fig::Bar_L} by definition. For all values above and right of the red curve, in each attack cycle, a transaction in the first block of $H$-chain is at risk of getting reversed in the sense that it is either not going to be confirmed at all or worse, it will be confirmed and then replaced with a conflicting transaction at no-cost to an adversary. The associated risks are provided in \figref{fig::btc_move} and \figref{fig::btc_double_sp}, respectively. As it is clear from Theorem~\ref{thm::db_sp}, $P(\textbf{Double-Spending})$ is decreasing in $\gamma$ due to the fact that even if the attack cycles end successfully, $\gamma$ increases the probability of \textbf{Move-Funds} event as the prefix of the $H$-chain switches to $\mathsf{B'}$ before $\mathsf{B}$ is confirmed because the adversary matches each honest block before $A=L$. Hence, for a given $\alpha$ as $\gamma$ increases, both  $P(\textbf{Double-Spending})$ and $P(\textbf{Service})$ decrease, whereas $P(\textbf{Move-Funds})$ increases. Further, as mentioned earlier, $P(\textbf{Service})>\beta$ since the first honest block $\mathsf{B}$ becomes part of the offset chain as soon as it is mined w.p. $\beta$. If one were to consider \textit{pre-mining} strategies, $P(\textbf{Double-Spending})$ could be increased at the expense of $P(\textbf{Service})$, hence security threats can be even more imminent than we display here.

\begin{figure}[t!]
\captionsetup[subfigure]{aboveskip=1pt,belowskip=9pt}
     \centering
     \begin{subfigure}[b]{0.48\columnwidth}
         \centering
         \includegraphics[width=\textwidth]{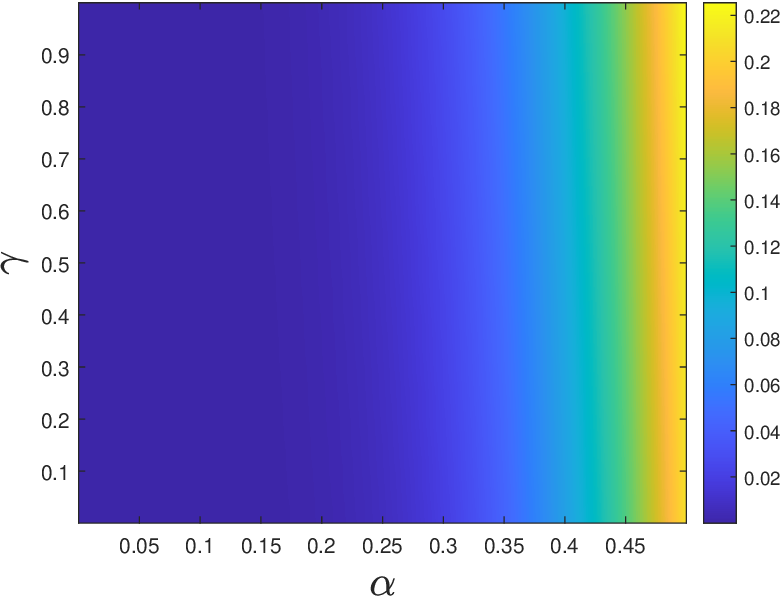}
         \caption{P(\textbf{Double-Spending})}
         \label{fig::btc_double_sp_stealth}
     \end{subfigure}
     \hfill
     \begin{subfigure}[b]{0.48\columnwidth}
         \centering
         \includegraphics[width=\textwidth]{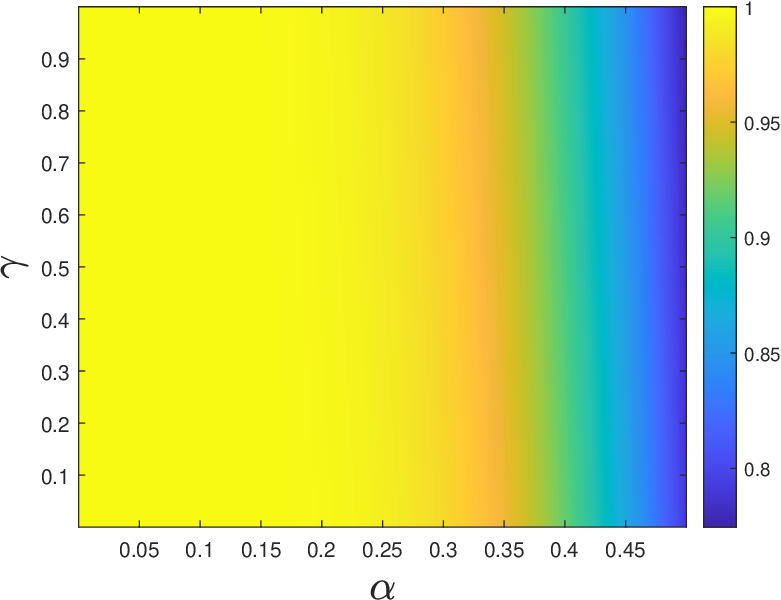}
         \caption{P(\textbf{Service})}
         \label{fig::btc_service_stealth}
     \end{subfigure}
     \hfill
     \begin{subfigure}[b]{0.48\columnwidth}
         \centering
         \includegraphics[width=\textwidth]{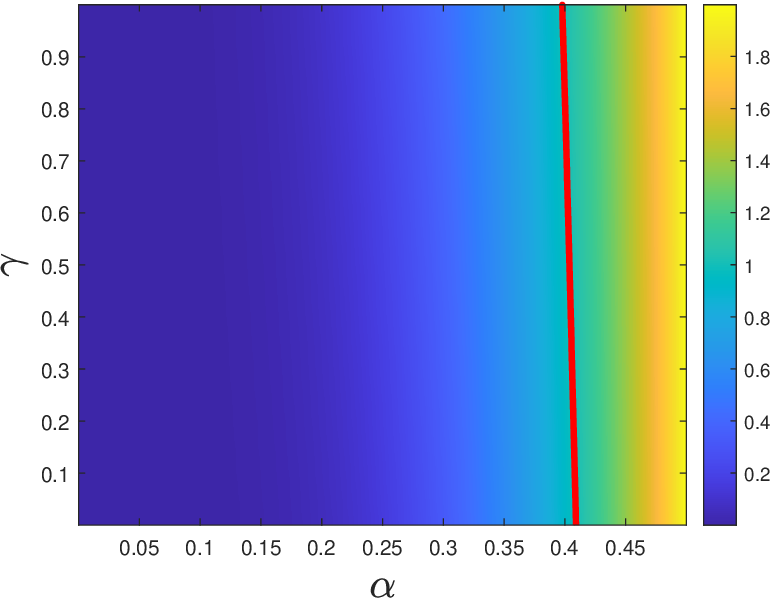}
         \caption{$\frac{\sigma_{k+1}}{\alpha}$}
         \label{fig::btc_revenue_scaled_stealth}
     \end{subfigure}
    \caption{$(k+1)$-stealth mining with $k=6$.}
	\label{fig::k_plus_1_probs_stealth}
\end{figure}

Next, in \figref{fig::k_plus_1_probs_stealth}, we provide the probabilities of the events \textbf{Double-Spending} and \textbf{Service} for each attack cycle of $(k+1)$-stealth mining with $k=6$ and the normalized revenue ratio. Again, at any point above and right of the red curve, a transaction in the first block of $H$-chain is at the risk of getting replaced by a conflicting transaction after being confirmed at no-cost to an adversary. The associated risk at each attack cycle is shown in \figref{fig::btc_double_sp_stealth}. We note that compared to stubborn mining, in stealth mining, for a given $\alpha$, increasing $\gamma$ increases the probability of \textbf{Double-Spending} as we were aiming. However, in stealth mining, for any given $\alpha$ and $\gamma$, $P(\textbf{Service})$ is larger compared to stubborn mining as an attack cycle cannot end with \textbf{Move-Funds} event in stealth mining. Essentially, in addition to the revenue decrease compared to the $L$-stubborn mining, this is the other cost of performing the attack by stealth and increasing $P(\textbf{Double-Spending})$.

For $(k+1)$-stubborn mining, the double-spending risk for $k=6$, $\alpha=0.41$ and $\gamma=0$ is $0.092$. For $k=6$ and $\gamma=0$, as $\alpha$ approaches $0.5$, the double-spending risk approaches $0.21$. On the other hand, for $k=6$, $\alpha=0.41$ and $\gamma=1$, the double-spending risk is $0.002$, whereas service probability is $0.59$. For $\gamma=0$, the same values apply to $(k+1)$-stealth mining as well. On the other hand, for $(k+1)$-stealth mining with $k=6$, $\alpha=0.41$ and $\gamma=1$, the double-spending risk is increased to $0.108$, whereas service probability increases to $0.892$. As mentioned earlier, the cost is a cut in revenue ratio, which drops from $\frac{\rho_{7}}{\alpha}=1.639$ to $\frac{\sigma_7}{\alpha}=1.09$.

\begin{figure}[t!]
     \centering
     \begin{subfigure}[b]{0.48\columnwidth}
         \centering
         \includegraphics[width=\textwidth]{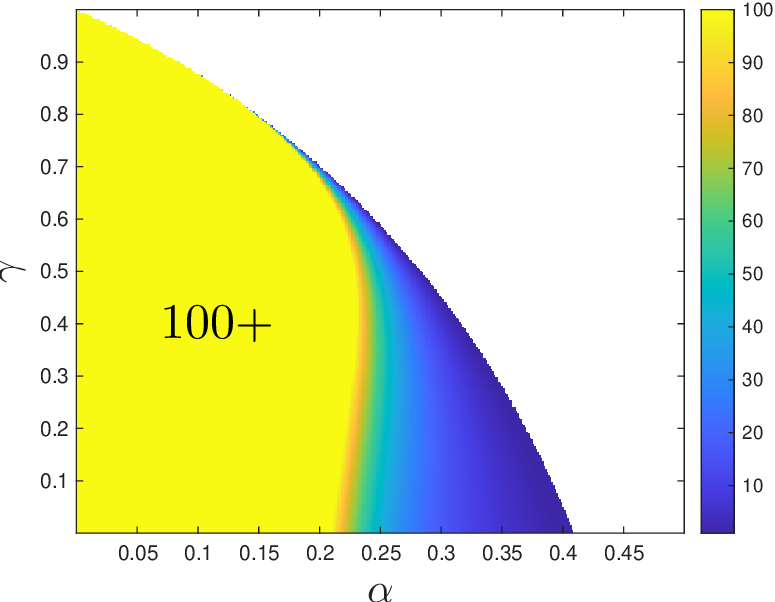}
         \caption{Stubborn mining}
         \label{fig::btc_least_R}
     \end{subfigure}
     \hfill
     \begin{subfigure}[b]{0.48\columnwidth}
         \centering
         \includegraphics[width=\textwidth]{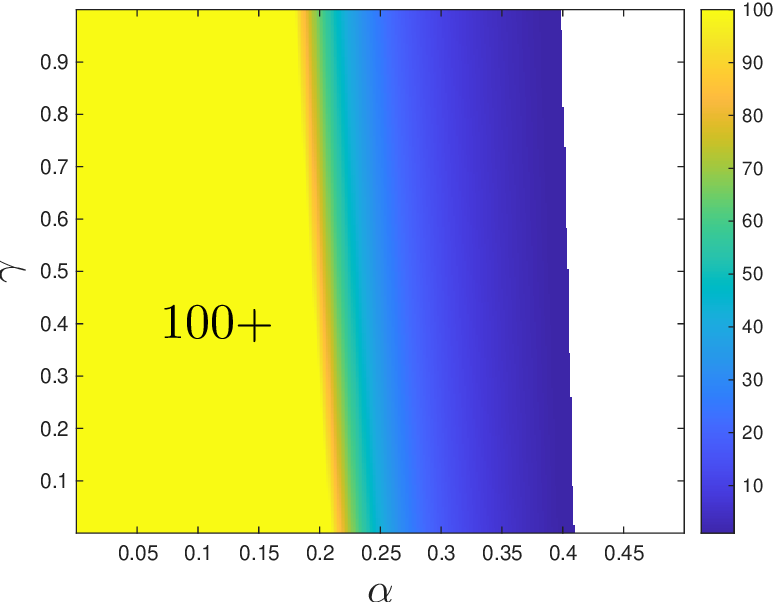}
         \caption{Stealth mining}
         \label{fig::btc_least_R_stealth}
     \end{subfigure}
    \caption{$R^*_{6}$ for Bitcoin}
	\label{fig::btc_least_Rs}
\end{figure}

In \figref{fig::btc_least_Rs}, we provide $R^*_{k}$ for each $\alpha$ and $\gamma$ value for Bitcoin, where $k=6$, for both stubborn mining and stealth mining attacks defined earlier. The white region is the region where $R^*_{k}=0$, \textit{i.e.}, the revenue ratio is already larger than honest mining. As the modification we provided for the attack resulted in larger $P(\textbf{Double-Spending})$ in stealth mining, $R^*_{k}$ value is quite low even for $\alpha$ and $\gamma$ values with $\frac{\sigma_{k+1}}{\alpha}<1$. On the other hand, it seems for $\alpha<0.2$, the double-spending rewards need to be extremely large for both of the attacks to be profitable.

\section{Conclusion}\label{sec::conclusion}
In this paper, we provided a rigorous and explicit stochastic analysis for the combination of the two celebrated attacks: double-spending and selfish mining. To do so, we first combined stubborn and selfish mining attacks, \textit{i.e.}, constructed a strategy where the attacker acts stubborn until its private branch reaches a certain length $L$, and then switches to act selfish. We provided an analysis of the optimal stubbornness to get the maximum revenue ratio for each parameter regime as well as the maximum stubbornness that is still more profitable than honest mining. Next, we connected the attack to double-spending, as $L > k$ implies double-spending. Thus, given $\alpha$ and $\gamma$, the maximum profitable stubbornness values determine the values of $k$ where $k$-confirmation rule is at risk of double-spending which comes at no-cost to the adversary. We also provided an exact formula for the double-spending probabilities (\textit{i.e.}, the risk) associated with each attack cycle. We provided the minimum double-spend value needed for an attack to be profitable in the regimes where the scheme is less profitable than honest mining. In order to conceal the  double-spending attack as well as to increase the associated successful double-spending probabilities, we modified the attack that does not \textit{match} every honest block during the stubborn phase. We provided all the relevant analysis for this modified attack as well.

\subsection{Discussion}
Notice that, in the literature, the revenue loss of the adversary is usually ignored when considering the double-spending attacks except for the MDP models of \cite{sompolinsky2016bitcoin,on_the_sec_pow_gervais} where a stochastic explicit analysis is missing and MDPs are used as black boxes to solve the problem. On the other hand, studies such as \cite{profitable_double_sp} focus on the double-spend value and only consider the usual double-spending attack structure of \cite{rosenfeld2014analysis}. The attack cycles we describe and analyze in this paper allow an attacker to mount a sustained double-spending attack until it succeeds while protecting the attacker from revenue loss in terms of the coinbase rewards in the long run by combining the attack with selfish mining.  Here, it is worth noting the severity of the selfish mining and double-spending attacks in general. The selfish mining threatens the fairness of the incentives. A selfish mining attack decreases the revenue and revenue ratio of all honest miners. Under the attack, if the honest mining is not sustainable anymore when the mining costs are taken into account, honest miners leave the system, increasing attackers fraction of hashrate and destabilizing the entire system. Even if the honest mining is sustainable despite the unfair reward distribution, the rational miners  can profit more if they join the adversary, destabilizing the entire system again. 

On the other hand, the severity of the implications of double-spending attacks  depends on how frequent it happens and the value of the transaction at question. It is no doubt that in high-value blockchain systems, double-spending a valuable asset is catastrophic. To prevent such a situation, confirmation latencies are calibrated. For example, with the $k=6$, \textit{i.e.}, the rule-of-the-thumb in Bitcoin, a successful double-spending attack probability for $\alpha=0.1$ is around $10^{-3}$ under $\Delta$-synchrony with $\Delta=10s$, despite the relatively low fork rate compared to for example Litecoin or Ethereum (back when it was PoW) \cite{our-sec-lat-extended}. If we consider a selfish mining attacker with $\alpha=0.3$, the success probability rises above $0.25$. Even if the confirmation latencies are calibrated from $k=6$ to $k=50$ for a high value transaction, the success probability is still around $10^{-4}$. However, these probabilities are obtained assuming worst case scenarios (in the favor of the adversary) in the state-of-the-art models of \cite{our-sec-lat-extended,cao2023tradeoff} and do not account for the coinbase reward losses, which potentially disincentivizes the attacker. Our model provides much more practical and realistic model for the attacker since in our model, the attacker is trying to keep an eye on the honest branch and releases its chain if necessary in order not to lose the coinbase rewards.

Even if we limit our focus on selfish mining, our modeling can be useful for analysis of different problems analytically. In our paper, we already solved some, e.g., parameterizing the mining strategies with $L$ and providing a method to find $L^*$ that achieves the optimal revenue ratio among the $L$-stubborn mining strategies. Another problem we solved was to parameterize the mining strategies that do not expose the selfish mining attack and to provide a method to find the one that achieves the optimal revenue ratio among these stealth mining strategies. Further, our method is a good analytical substitute for the OSM MDP. For example, it can be directly used in the analysis of revenue per chain progress for selfish mining attacks considered in \cite{time_average_selfish_mining}.

From the perspective of consistency of the protocol, $L$-stubborn mining and $S$-stealth mining attacks do not aim to replace the offset chain. The offset chain is extended at the end of every attack cycle and in the long run asymptotically equals the longest chain. At every attack cycle, the aim of the attacker is to cause two parallel forks extending the offset chain, \textit{i.e.}, $A$-chain and $H$-chain. The fork in turn allows the adversary to get a chance at replacing the honest blocks of $H$-chain, stealing their revenue and double-spending. The fork that loses this competition at the end of each attack cyle essentially joins the orphan blocks of the blockchain tree. Within this attack model, the block at the tip of the offset chain can be seen as a \textit{Nakamoto block} \cite{nakamoto-always-wins}, \textit{i.e.}, it is permanently confirmed since both the attacker and the honest miner accept them. On the other hand, double-spending happens because the honest miners and merchants accept the blocks according to the $k$-confirmation rule which is a heuristic and are not able to pinpoint the tip of the offset chain, especially if the attacker is using $S$-stealth mining attack to conceal its $A$-chain by avoiding the \textit{matching} action. 

\subsection{Extensions}\label{sec::extensions}
There are two possible extensions to our paper that can further increase the action space and combining the two would eventually allow all possible state-space considered in OSM MDP. First, note that, we define the attack cycles in a way such that they are i.i.d. This is a design choice which allows us to know whether $\mathsf{B}$ is confirmed or not and use Doob's optional stopping theorem in our analysis as well as the techniques provided in Appendix~\ref{sec::app-l-stub} and Appendix~\ref{sec::app-s-steal}. In this paper, each time an attack cycle starts, we have $A=0$ and $H=0$, since we essentially keep track of the length of the $A$-chain, even if the $H$-chain has some prefix consisting of adversarial blocks of the $A$-chain. On the other hand, one could consider the situations where the duration of attack cycles are defined with respect to the change in the prefix of the $H$-chain, \textit{i.e.}, restricting $H$-chain to consist of the honest blocks exclusively, which is essentially considered in the OSM MDP. This change not only complicates the rigorous analysis and requires us to keep a parameter that tracks the number of honest blocks, but also requires us to keep track of the depth of the $\mathsf{B}$ and $\mathsf{B'}$, hence our design choice is to avoid this. The other extension is more obvious, which is to allow trail stubbornness, \textit{i.e.}, to allow the adversary to continue the attack even if it falls behind. By considering an extra parameter for the upper bound on the deficit of the $A$-chain compared to the $H$-chain, our analysis can be redone, however, this would also complicate the analysis provided in Appendix~\ref{sec::app-l-stub} and Appendix~\ref{sec::app-s-steal}. Note that, increasing the action space for the attacker increases the maximum revenue ratio that can be obtained. However, based on our simulations and comparison with the maximum revenue ratios obtained by the OSM MDP for several $(\alpha,\gamma)$ parameters, the difference is negligible.

\appendices
\section{Stochastic Tools for Analysis}\label{sec::preliminaries}

\subsection{Catalan Numbers and Dyck Words}
Let $S$ be a string consisting of $X$'s and $Y$'s. If no prefix of $S$ has more $Y$'s than $X$'s, then we call $S$ a Pre-Dyck word. If a Pre-Dyck word has equal number of $X$'s and $Y$'s, it is called a Dyck word. Any Pre-Dyck word can be converted to a Dyck word by appending enough number of $Y$'s. The total number of Pre-Dyck words that contain $n$ $X$'s and $m$ $Y$'s is denoted as $P[n,m]$. The number of Dyck words of length $2n$, \textit{i.e.}, $P[n,n]$, is called Catalan number $C_n$ \cite{catalan-numbers-book}, and satisfies
\begin{align}
    C_n=\frac{1}{n+1} {2n \choose n}\label{eq::catalan_formula}.
\end{align}
Catalan numbers also satisfy the following recursion
\begin{align}
    C_n=\frac{2(2n-1)}{n+1}C_{n-1},
\end{align}
with $C_0=1$. The generating function of the Catalan numbers is
\begin{align}
C(x)=\sum_{n\geq 0}C_n x^{n}=\frac{2}{1+\sqrt{1-4x}}.\label{eq::catalan_generator} 
\end{align}  
Pre-Dyck words satisfy the recursion
\begin{align}
    P[n,m]=P[n,m-1]+P[n-1,m] \label{eq::Pre-Dyck-recursion}.
\end{align}
Clearly $P[n,m]=0$ if $m>n$ and it is easy to verify that $P[n,0]=1$ and $P[n,1]=n$. Thus, \eqref{eq::Pre-Dyck-recursion} gives a straightforward formula for finding $P[n,m]$ in $\mathcal{O}(nm)$.

\subsection{Bailey's Number and Bertrand's Ballot Problem}
Although it is easy to find a formula for $P[n,m]$ by converting \eqref{eq::Pre-Dyck-recursion} to frequency domain using generating functions, one could draw an analogy to the well-known Bertrand's ballot problem \cite{ballot-problem}, which asks the probability that a candidate A, who received $n$ votes, was strictly ahead of a candidate B, who received $m$ votes, throughout an election. By slightly modifying Andre's solution to the ballot problem \cite{andre-ballot-solution}, which can be used to count the number of paths from $(a,b)$ to $(c,d)$ above and below $y=x$, to allow ties, we find
\begin{align}
    P[n,m]=\frac{n-m+1}{n+m+1} {n+m+1 \choose n+1}. \label{eq::bailey}
\end{align}
The elements of the matrix $P$ constitute Catalan's triangle. Note that, if a broad range of $P[n,m]$ is needed, it is better to start from $P[1,0]=1$ and fill the matrix $P$ via \eqref{eq::Pre-Dyck-recursion}.  The elements of $P$ matrix appear in \cite{baileys_number} where Bailey investigates the sequences that contain $n$ $1$'s and $m$ $-1$'s such that the running sum is always non-negative. Bailey proved that the number of such sequences is equal to $P[n,m]$ provided in \eqref{eq::bailey}.

\subsection{Wald's Identity and Doob's Optional Stopping Time }\label{subsec::walds-identity}
Consider a Pre-Dyck word with $L$ $X$'s and $m<L$ $Y$'s. Append a random sequence of $X$'s (w.p. $\alpha$) and $Y$'s (w.p. $1-\alpha$) until the first time the number of $Y$'s becomes 1 less than the number of $X$'s. As this is a stopping time, by Wald's identity, the expected length of such a sequence is $\frac{L-m-1}{1-2\alpha}$. As $L-m-1$ $Y$'s are required to reduce the gap to $1$, the rest of the characters in the string are split evenly between $X$ and $Y$. Thus, the final sequence contains $L+\frac{(L-m-1)\alpha}{1-2\alpha}$ $X$'s in expectation.

\begin{table*}[t!]
\centering

\begin{tabular}{|>{\centering\arraybackslash}m{4.8cm}|>{\centering\arraybackslash}m{4.8cm}|>{\centering\arraybackslash}m{2.5cm}|>{\centering\arraybackslash}m{2.3cm}|}
\hline
\textbf{State $\times$ Action} & \textbf{State} & \textbf{Probability} & \textbf{Reward} \\
\hline
$(a, h, \cdot), \text{adopt}$ & $(1, 0, \text{irrelevant})$ & $\alpha$ & $(0, h)$ \\
\cline{2-4}
                              & $(0, 1, \text{irrelevant})$ & $1 - \alpha$ & $(0, h)$ \\
\hline
$(a, h, \cdot), \text{override}^\dagger$ & $(a - h, 0, \text{irrelevant})$ & $\alpha$ & $(h + 1, 0)$ \\
\cline{2-4}
                              & $(a - h - 1, 1, \text{relevant})$ & $1 - \alpha$ & $(h + 1, 0)$ \\
\hline
$(a, h, \text{irrelevant}), \text{wait}$ & $(a + 1, h, \text{irrelevant})$ & $\alpha$ & $(0, 0)$ \\\cline{2-4}
$(a, h, \text{relevant}), \text{wait}$ & $(a, h+1, \text{relevant})$ & $1 - \alpha$ & $(0, 0)$ \\
\hline
$(a, h, \text{active}), \text{wait}$ & $(a + 1, h, \text{active})$ & $\alpha$ & $(0, 0)$ \\
\cline{2-4}
$(a, h, \text{relevant}), \text{match}^\ddagger$ & $(a - h, 1, \text{relevant})$ & $\gamma \cdot (1 - \alpha)$ & $(h, 0)$ \\
\cline{2-4}
                              & $(a, h + 1, \text{relevant})$ & $(1 - \gamma) \cdot (1 - \alpha)$ & $(0, 0)$ \\
\hline
\end{tabular}

\vspace{2mm}
\raggedright
\noindent\textsuperscript{$\ddagger$} feasible only when $a \geq h$ \\
\textsuperscript{$\dagger$} feasible only when $a > h$
\caption{State, action, and transition of OSM MDP \cite{optimal-selfish}}\label{table::osm-mdp}
\end{table*}

\section{Selfish Mining \cite{selfish-mining}, OSM MDP \cite{optimal-selfish} and Our Model}\label{sec::app-existing}
Here, we restate and summarize the important bits of the selfish mining attack of \cite{selfish-mining} as well as the Optimal Selfish Mining (OSM) MDP model introduced in \cite{optimal-selfish} to find the $\epsilon$-optimal selfish mining strategies. Then, we represent the state and action spaces for $L$-stubborn mining and $S$-stealth mining, which serve as formal descriptions of the strategies introduced in Section~\ref{sec::l-stub} and Section~\ref{sec::s-steal} for the sake of completeness.

\subsection{Selfish Mining of Eyal and Sirer \cite{selfish-mining}}
In essence, the original attack introduced in \cite{selfish-mining} works as follows. Stating from the same offset chain: 
\begin{enumerate}
    \item If the honest miners mine a block first, the attacker accepts the block and redefines the offset chain.
    \item\label{enum::step::1} If the adversary mines a block first, it hides it and tries to extend it. When the next block arrives:
    \begin{enumerate}
        \item \label{enum::step::2}  If the block is adversarial, the adversary has $2$ private blocks. It will keep mining privately until the difference between the public branch and private branch reduces to $1$, at which point it will release the private branch to override the public branch.
        \item If the block is honest, the adversary releases its private branch, \textit{i.e.}, \textit{matching}, resulting in a public fork of length $1$. At this point, one branch is adversarial and the other is honest, and $\gamma$ fraction of the honest miners prefer the adversarial branch to mine on. Whoever mines the next block, releases it immediately, and the offset chain is redefined.
    \end{enumerate}
\end{enumerate}
Notice that the attacker avoids the release of its block at Step~\ref{enum::step::1} initially. Our idea is to extend this further even in Step~\ref{enum::step::2} if the adversarial branch is not long enough, \textit{i.e.}, until the private branch reaches length $L$.  This is because, an override action means that the attacker is releasing one additional block to convince others that its released branch is longer than the honest branch, however, it also means the adversary is potentially losing some additional revenue. 
 
\subsection{OSM MDP of \cite{optimal-selfish}}
We start by stating the state space of the OSM MDP. A state is essentially of the $3$-tuple form $(a,h,fork)$, where $a$ and $h$ denote the lengths of the adversarial private and honest public branches, respectively. The last field $fork$ stores the information about the feasibility of the \textit{match} action. If $fork=relevant$, it means the adversary can take the action whereas if $fork\neq relevant$, it cannot. On the other hand, if a match action is taken recently, $fork=active$. We summarize the state action space and the transitions in Table~\ref{table::osm-mdp}, taken directly from \cite{optimal-selfish}.

Letting $r^t=(r^t_1,r^t_2)$ denote the reward for the transitions between the states according to the action taken at step $t$, we get the adversarial revenue ratio as
\begin{align}
    \rho&=\mathbb{E}\left[\lim_{T\to\infty}\frac{\sum_{t=1}^{T}r^t_1}{{\sum_{t=1}^{T}r^t_1 + \sum_{t=1}^{T}r^t_2}}\right].\label{eq::mdp_rew_ratio}
\end{align}
Next, define $w_\rho(x,y)$ as 
\begin{align}
    w_\rho(x,y)=(1-\rho)x-\rho y.
\end{align} 
Now, for any policy $\pi$, if \eqref{eq::mdp_rew_ratio} is rearranged,  we get
\begin{align}v^\pi_\rho&=\mathbb{E}\left[\lim_{T\rightarrow\infty}\frac{1}{T}\sum_{t=1}^{T} w_\rho(r^t_1,r^t_2)\right].
\end{align}
If a policy $\pi^*$ maximizes the revenue ratio and results in $\rho^*$, then $v^{\pi^*}_{\rho^*}=0$. Hence, \cite{optimal-selfish} uses the monotonicity of $w_\rho(x,y)$ in $\rho$ to find such an optimal policy. For further details about the OSM MDP formulation and its analysis, we refer the reader to \cite{optimal-selfish}.

\subsection{State$\times$Action Space for $L$-Stubborn and $S$-Stealth Mining}\label{app::state_action_stub_stealth}
\subsubsection{$L$-Stubborn Mining}
To describe the state space for $L$-stubborn mining, we assume a common block agreed by honest and adversarial miners (either \textit{adopt} or \textit{override} action is taken most recently), and the adversary starts a new attack cycle. Then, the state space for each $L$-stubborn mining attack cycle can be succinctly expressed as $(A,H,C,fork)$ where, starting from the common block (excluding), $A$ is the length of the adversarial branch  and $H$ is the length of the honest branch and $C$ is the length of the common prefix between the adversarial and honest branches. The field $fork$ serves the same purpose as in OSM MDP, \textit{i.e.}, if the last mined block is honest, $fork=relevant$, which implies that the adversary can take the match action whereas $fork=active$ means that a match action is taken recently and $\gamma$ fraction of honest miners are mining on a prefix of adversarial branch of the same length as the honest branch. Under this model, in Table~\ref{table::l_stub}, we summarize the state action space and the transitions of $L$-stubborn mining.

\begin{table*}[h!]\footnotesize
\centering

\begin{tabular}{|>{\centering\arraybackslash}m{4.8cm}|>{\centering\arraybackslash}m{4.8cm}|>{\centering\arraybackslash}m{2.5cm}|>{\centering\arraybackslash}m{2.3cm}|}
\hline
\textbf{State $\times$ Action} & \textbf{State} & \textbf{Probability} & \textbf{Reward} \\
\hline
$(A, H, C, \cdot), \text{adopt}^\dagger$ & $(1, 0,0, \text{irrelevant})$ & $\alpha$ & $(C,H-C)$ \\
\cline{2-4}
& $(0, 1,0, \text{irrelevant})$ & $1 - \alpha$ & $(C, H-C)$ \\
\hline
$(A, H,C, \cdot), \text{override}^\ddagger$ & $(1, 0, 0,\text{irrelevant})$ & $\alpha$ & $(A, 0)$ \\
\cline{2-4}
& $(0, 1,0, \text{irrelevant})$ & $1 - \alpha$ & $(A, 0)$ \\
\hline
$(A, H,C, \text{irrelevant}), \text{wait}$ & $(A + 1, H,C, \text{irrelevant})$ & $\alpha$ & $(0, 0)$ \\
\cline{2-4}
 & $(A, H+1,C, \text{relevant})$ & $1 - \alpha$ & $(0, 0)$ \\
\hline
$(A, H,C, \text{active}), \text{wait}$ & $(A+1, H,C, \text{active})$ & $\alpha$ & $(0, 0)$ \\
\cline{2-4}
$(A, H, C, \text{relevant}), \text{match}^\star$ & $(A, H+1,H, \text{relevant})$ & $\gamma \cdot (1 - \alpha)$ & $(0, 0)$ \\
\cline{2-4}
& $(A, H + 1,C, \text{relevant})$ & $(1 - \gamma) \cdot (1 - \alpha)$ & $(0, 0)$ \\
\hline
\end{tabular}
\vspace{2mm}

\raggedright
\noindent\textsuperscript{$\dagger$} happens whenever $A+1=H$ \\
\textsuperscript{$\ddagger$} happens whenever $A=H+1\geq L$\\
\textsuperscript{$\star$} happens whenever $L>A\geq H$
\caption{State, action, and transitions of $L$-Stubborn Mining}\label{table::l_stub}
\end{table*}
Notice, in $L$-stubborn mining each state allows a unique action to be taken as presented in Table~\ref{table::l_stub} as opposed to OSM MDP where the goal is to find the optimal action among multiple options for each state. For example, here, whenever $A+1=H$, the action \textit{adopt} is executed, \textit{i.e.}, no-trail mining is allowed whereas \textit{override} action is taken whenever $A=H+1\geq L$. Similarly, whenever $fork=relevant$ and $L>A\geq H$, the adversary \textit{matches} the recently mined honest block. For all the rest of the cases, \textit{wait} action is taken. Our focus in the analysis is to express the final revenue ratio in terms of $L$ as well as to maximize the revenue ratio with respect to $L$ which essentially changes when each action is taken. Although changing $L$ essentially gives multiple strategies to pick from, this does not cover all the possible strategies as in OSM MDP, hence the final optimal $\rho_{L^*}$ is still below $\rho_{OSM}$.

An important observation from Table~\ref{table::l_stub} is that, the value $C$ is only affected by the value of $H$ in a Markovian manner as represented in the last two rows of Table~\ref{table::l_stub}. Further, none of the actions depend on $C$ field in the state space and $C$ only affects the reward if an \textit{adopt} action is taken. This implies that during an $L$-stubborn attack cycle, given $H$, we have
\begin{align}
    C = \begin{cases}
        H - 1 - i,  &\text{w.p. } (1 - \gamma)^i \gamma, \quad  0 \leq i < H - 1, \\
        0, & \text{w.p. } (1 - \gamma)^{H - 1}.
    \end{cases}
\end{align}
This explains why we treat $C$ as a separate state in the analysis throughout the paper and only introduce here for the sake of completeness. Similarly, since every honest block that is mined on a height where the adversary already has a private block, is \textit{matched} in $L$-stubborn mining, we did not need to introduce $fork$ field in the main body of our paper.
\subsubsection{$S$-Stealth Mining}
In $S$-stealth mining, \textit{match} action only happens when in the previous state $S-1=A=H+1$ and a new honest block is mined such that the current state is $S-1=A=H$ at which point the adversary releases its $A$-chain to compete with the honest branch of the same length. As a result, we do not need $C$ and $fork$ field in this model and the state $\times$ action space is given in Table~\ref{table::s_stealth}.
\begin{table*}[h!]\footnotesize
\centering

\begin{tabular}{|>{\centering\arraybackslash}m{4.8cm}|>{\centering\arraybackslash}m{4.8cm}|>{\centering\arraybackslash}m{2.5cm}|>{\centering\arraybackslash}m{2.3cm}|}
\hline
\textbf{State $\times$ Action} & \textbf{State} & \textbf{Probability} & \textbf{Reward} \\
\hline
$(A, H), \text{adopt}^\dagger$ & $(1, 0)$ & $\alpha$ & $(0,H)$ \\
\cline{2-4}
& $(0, 1)$ & $1 - \alpha$ & $(0, H)$ \\
\hline
$(A, H), \text{override}^\ddagger$ & $(1, 0)$ & $\alpha$ & $(A, 0)$ \\
\cline{2-4}
& $(0, 1)$ & $1 - \alpha$ & $(A, 0)$ \\
\hline
$(A, H), \text{wait}$ & $(A + 1, H)$ & $\alpha$ & $(0, 0)$ \\
\cline{2-4}
 & $(A, H+1)$ & $1 - \alpha$ & $(0, 0)$ \\
\hline
$(A, H), \text{match}^\star$ & $(0, 0)$ & $\alpha$ & $(A+1, 0)$ \\
\cline{2-4}
 & $(0, 0)$ & $\gamma \cdot (1 - \alpha)$ & $(A, 1)$ \\
\cline{2-4}
& $(0,0)$ & $(1 - \gamma) \cdot (1 - \alpha)$ & $(0, H+1)$ \\
\hline
\end{tabular}

\vspace{2mm}
\raggedright
\noindent\textsuperscript{$\dagger$} happens whenever $A+1=H$ \\
\textsuperscript{$\ddagger$} happens whenever $A=H+1\geq S$\\
\textsuperscript{$\star$} happens whenever $S-1=A= H$
\caption{State, action, and transitions of $S$-Stealth Mining}\label{table::s_stealth}
\end{table*}

\section{Proof of Theorem~\ref{thm::quasiconcavity} and Algorithm~\ref{alg::cap}} \label{sec::app-l-stub}
As in \cite{optimal-selfish}, we define $w_\rho(x,y)$ as 
\begin{align}
    w_\rho(x,y)=(1-\rho)x-\rho y.
\end{align}
Let $x_i$ and $y_i$ be the number of adversarial and honest blocks that make it into the offset chain at the end of an attack cycle $i$ of $L$-stubborn mining strategy. Clearly, the revenue ratio in the long run is 
\begin{align}
    \rho&=\lim_{T \to \infty}\frac{\sum_{i=1}^{T}x_i}{{\sum_{i=1}^{T}x_i + \sum_{i=1}^{T}y_i}}\\
    &=\frac{\mathbb{E}[x_i]}{\mathbb{E}[x_i+y_i]},\label{eq::stopping_thm_res}
\end{align} 
where \eqref{eq::stopping_thm_res} follows from the strong law of large numbers and i.i.d.~attack cycles. Moving terms, we get $w_\rho(\mathbb{E}[x_i],\mathbb{E}[y_i])=0$, which is similar to the linearization of OSM MDP \cite{optimal-selfish}. 

Notice that, on average, $\rho$ fraction of the rewards belongs to the adversary. Hence, $w_\rho(x_i,y_i)$ is essentially a function that indicates how much the reward distribution of the cycle $i$ deviates from the average. Using this interpretation, we show that, when the strategies are restricted to $L$-stubborn mining strategies, the analysis can be narrowed down to a specific case of a single cycle $(A=H+1)$, where a decision about override has to be made or delayed. Here, our goal is to prove that, we can only check the cases of $A=H+1$, instead of implementing an MDP directly.

Assume, the current cycle $i$ of the $L$-stubborn mining strategy has $A=x$ and $H=y$ (cycle has not ended yet necessarily). Let $w^{(L)}_\rho(x,y)$ denote the expected value of $w_\rho$ associated with the cycle $i$ given $A=x$ and $H=y$ and the attacker employs $L$-stubborn mining strategy, \textit{i.e.},
\begin{align}
    &w^{(L)}_\rho(x,y)\nonumber\\&=w_\rho(\mathbb{E}^{(L)}[x_i|A=x,H=y],\mathbb{E}^{(L)}[y_i|A=x,H=y])\\
    &=(1-\rho)\mathbb{E}^{(L)}[x_i|A=x,H=y]-\rho\mathbb{E}^{(L)}[y_i|A=x,H=y],
\end{align}
where the superscript $(L)$ means that the adversary follows $L$-stubborn mining strategy. For example, when $A\geq L$ and $H=A-1$, $w^{(L)}_\rho(A,A-1)=(1-\rho)A$ since the decision is to release the $A$-chain immediately. On the other hand, if $A+1=H\leq L$, where $H$-chain has $a$ adversarial blocks in its prefix, then $w^{(L)}_\rho(H-1,H)=(1-\rho)a-\rho(H-a)$ since the decision is to accept the $H$-chain and abort the attack cycle.

\begin{lemma}\label{lemma::bounds_strategies}
If $\rho_{L+1}<\rho_L$, then
\begin{align}
    &w^{(L+1)}_{\rho_{L}}(L,L-1) < w^{(L+1)}_{\rho_{L+1}}(L,L-1) \nonumber\\&< w^{(L)}_{\rho_{L}}(L,L-1) <  w^{(L)}_{\rho_{L+1}}(L,L-1). \label{eq:lemma_strategies_relations}
\end{align}
\end{lemma}

\begin{Proof}
    Let $q_{L}$ denote the probability that a cycle passes through $A=H+1=L$. Then,
\begin{align}
    0&=\lim_{T\to\infty}\frac{1}{T}\sum_{i=1}^{T}w^{(L)}_{\rho_{L}}(x_i,y_i)\\
    &=q_{L}w^{(L)}_{\rho_{L}}(L,L-1)+(1-q_{L})w^{(L)}_{\rho_{L}}(\overline{L,L-1}), \label{eq::a_strategy}
\end{align}
where $w^{(L)}_{\rho_{L}}(\overline{L,L-1})$ is the expected value of $w_\rho$ for $L$-stubborn mining strategy given $(\overline{L,L-1})$, \textit{i.e.}, the cycle does not pass through $A=H+1=L$. Next, for any $\rho\in [0,1]$
\begin{align}
    w^{(L)}_{\rho}(\overline{L,L-1})=w^{(L+1)}_{\rho}(\overline{L,L-1}),
\end{align}
which is due to the fact that both strategies have the same rewards for all possible arrival paths that do not pass through the event $A=H+1=L$ during a cycle. To see this, consider $4$-stubborn mining and $5$-stubborn mining, for example. All red crosses represented in \figref{fig::coordinate} for $4$-stubborn mining are also present in $5$-stubborn mining, \textit{i.e.}, the cycles have the same decisions as long as $A<4$. Similarly, all blue crosses except the blue cross at $(4,3)$ represented in \figref{fig::coordinate} for $4$-stubborn mining are also present in $5$-stubborn mining, \textit{i.e.}, the cycles have the same decisions as long as $A>4$. Thus, the only difference in rewards is, if a cycle passes through the point $(4,3)$.   

Further, if $\rho_{L+1}<\rho_L$, as $w_\rho$ is decreasing in $\rho$, we have
\begin{align}
    w^{(L)}_{\rho_{L}}(\overline{L,L-1})< w^{(L+1)}_{\rho_{L+1}}(\overline{L,L-1}).\label{eq::lvslplus1}
\end{align}
Thus, rewriting \eqref{eq::a_strategy} for $(L+1)$-stubborn mining strategy
\begin{align}
    0&=q_{L}w^{(L+1)}_{\rho_{L+1}}(L,L-1)+(1-q_{L})w^{(L+1)}_{\rho_{L+1}}(\overline{L,L-1})\\
    &> q_{L}w^{(L+1)}_{\rho_{L+1}}(L,L-1)+(1-q_{L})w^{(L)}_{\rho_{L}}(\overline{L,L-1})\label{eq::a_strategy_midstep}\\
    &=q_{L}\left(w^{(L+1)}_{\rho_{L+1}}(L,L-1)-w^{(L)}_{\rho_{L}}(L,L-1)\right), \label{eq::a_strategy_proved}
\end{align}
where \eqref{eq::a_strategy_midstep} follows from \eqref{eq::lvslplus1} and \eqref{eq::a_strategy_proved} follows from \eqref{eq::a_strategy}.
This proves the inequality in the middle of \eqref{eq:lemma_strategies_relations}. The other two inequalities follow from the fact that $w_\rho$ is decreasing in $\rho$.
\end{Proof}

\begin{corollary}\label{corollary::bounds_strategies}
    If $\rho_{L+1}>\rho_L$, then
\begin{align}
 &w^{(L+1)}_{\rho_{L}}(L,L-1) > w^{(L+1)}_{\rho_{L+1}}(L,L-1)\nonumber\\ &> w^{(L)}_{\rho_{L}}(L,L-1) >  w^{(L)}_{\rho_{L+1}}(L,L-1).\label{eq:cor_strategies_relations}
\end{align}
\end{corollary}

\begin{Proof}
    Simply change the direction of the inequalities from \eqref{eq::lvslplus1} onwards in the proof of Lemma \ref{lemma::bounds_strategies}.
\end{Proof}

\begin{lemma}\label{lemma::eq_strategies}
Given $A=H+1=L$, the expected result of $w_\sigma$ associated with $(L+1)$-stubborn mining strategy is related to $L$-stubborn mining strategy in the following way
    \begin{align}
    &w^{(L+1)}_{\rho}(L,L-1)\nonumber\\&=w^{(L)}_{\rho}(L,L-1)+\beta^2\left(\frac{1-{\rho}}{1-2\alpha}-\frac{1}{\gamma}+\frac{(1-\gamma)^{L+1}}{\gamma}\right).\label{eq::narrow-decision}
\end{align}
\end{lemma}

\begin{Proof}
    Clearly, if $L$-stubborn mining strategy is followed, we get $ w^{(L)}_{\rho}(L,L-1)=w_{\rho}(L,0)$. On the other hand, if $(L+1)$-stubborn mining strategy is followed, the decision is delayed and $w^{(L+1)}_{\rho}(L,L-1)$ depends on the arrival of the next blocks in the following way:
    \begin{enumerate}
    \item The next block is adversarial w.p. $\alpha$. In this case, the adversarial advantage $A-H=2$ implies that the decision will be made when the advantage reduces to $1$, which happens after $\frac{1}{1-2\alpha}$ block arrivals in expectation and $\frac{\alpha}{1-2\alpha}$ of those blocks will be adversarial. Hence, w.p. $\alpha$ we get $w_{\rho}(L+1+\frac{\alpha}{1-2\alpha},0)$.
    \item The next block is honest followed by an adversarial block w.p. $\beta\alpha$. As a result, we get $w_{\rho}(L+1,0)$ w.p. $\beta\alpha$.
    \item The next two blocks are honest w.p. $\beta^2$. In this situation, $w_{\rho}$ depends on how many adversarial blocks are contained in the prefix of the honest block at height $L+1$:
    \begin{enumerate}
        \item The prefix contains $L-i>0$ adversarial blocks w.p. $\gamma(1-\gamma)^{i}$ and we have $w_{\rho}(L-i,i+1)$.
        \item The prefix does not contain any adversarial block w.p. $(1-\gamma)^L$ and we have $w_{\rho}(0,L+1)$.
    \end{enumerate}
\end{enumerate}
Combining the cases above in expectation, for $(L+1)$-stubborn mining strategy, we obtain
\begin{align}
    &w^{(L+1)}_{\rho}(L,L-1)=\alpha w_{\rho}(L+1+\frac{\alpha}{1-2\alpha},0)\nonumber\\&+\beta\alpha w_{\rho}(L+1,0)+\beta^2(1-\gamma)^L w_{\rho}(0,L+1)\nonumber\\&+\sum_{i=0}^{L-1}\beta^2\gamma(1-\gamma)^i w_{\rho}(L-i,i+1),
\end{align}
which results in \eqref{eq::narrow-decision}.
\end{Proof} 

Notice what we did in this sequence of lemmas: In Lemma~\ref{lemma::bounds_strategies}, given the current attack cycle has $A=L=H+1$, we established a relation between the expected value of $w_{\rho_{L}}$ and $w_{\rho_{L+1}}$ for $L$-stubborn mining strategy and $L+1$-stubborn mining strategy assuming $\rho_{L}>\rho_{L+1}$ as well as the opposite situation in Corollary~\ref{corollary::bounds_strategies}. On the other hand, in Lemma~\ref{lemma::eq_strategies}, we established a direct relation between the expected value of $w_{\rho}$ for $L$-stubborn mining strategy and $L+1$-stubborn mining strategy. Hence, to prove Theorem~\ref{thm::quasiconcavity}, all that is left is to check when these relationships hold.

Let $u$ and $v_a$ be,
\begin{align}
    u=(1-\gamma), \quad\quad 
    v_{a}=\frac{1}{u}\left(1-\gamma\frac{1-\rho_{a}}{1-2\alpha}\right).
\end{align}
If $\rho_{L}\geq\rho_{L+1}$, picking $\rho=\rho_{L}$ and $\rho=\rho_{L+1}$ in \eqref{eq::narrow-decision} implies
\begin{align}
     v_{L} \geq v_{L+1} &\geq u^L, \label{eq::lemmacombine}
\end{align}
due to Lemma~\ref{lemma::bounds_strategies}.
 
Next, assume $\rho_{L+1}<\rho_{L+2}$ for the sake of contradiction. Rewriting \eqref{eq::narrow-decision} and \eqref{eq:cor_strategies_relations} for the relation between $(L+1)$ and $(L+2)$-stubborn mining and picking $\rho=\rho_{L+1}$ in \eqref{eq::narrow-decision}, we get
\begin{align}
     u^{L+1}> v_{L+1},
\end{align}
which creates a contradiction with \eqref{eq::lemmacombine} since $u^L> u^{L+1}$. The rest is just applying the recursion and the same proof applies for both claims in the theorem statement. 

\begin{proposition}\label{prop::self}
    Algorithm~\ref{alg::cap} returns $L^*$.
\end{proposition}

\begin{Proof}
    First, assume  $L^*=\infty$, then clearly $L^{(1)}=\infty$ and $\rho_{L+1}>\rho_{L}$ implies
    \begin{align}
        u^{L}> v_{L},
    \end{align}
    for all $L$. Taking limit of both sides implies $0\geq v_{L^{(1)}}$, thus, algorithm returns $\infty$.

    Next, assume $L^*$ is finite. Then, for all $L>L^*$, $\rho_L>\rho_{L+1}$, which implies
    \begin{align}
        u^{L}< v_{L}.
    \end{align}
    Taking limit of both sides further implies that
    \begin{align}
        0<v_{\infty} \leq \max\{v_1,v_2,v_{\infty}\},
    \end{align}
    \textit{i.e.}, the algorithm does not return infinity. Next assume $L^*=1$, then clearly $L^{(1)}=1$ and the algorithm returns $1$. 
    
    If $1<L^*<\infty$, then clearly $L^{(1)}=2$ or $L^{(1)}=\infty$, and the algorithm enters the while loop. From Lemma~\ref{lemma::bounds_strategies},  Lemma~\ref{lemma::eq_strategies} and Corollary~\ref{corollary::bounds_strategies}, if $\rho_L > \rho_{L+1}$
    \begin{align}
        \frac{\log v_{L}}{\log u}< \frac{\log v_{L+1}}{\log u}< L, \label{eq::conv_up}
    \end{align}
    and if $\rho_L > \rho_{L-1}$
    \begin{align}
        \frac{\log v_{L-1}}{\log u}> \frac{\log v_{L}}{\log u}> L-1. \label{eq::conv_low}
    \end{align}
    Further, note that $v_{L^{*}}\geq v_a$ for any $a\in \mathbb{Z}$. Assuming $L^*$ is finite, \eqref{eq::conv_low} implies, for all $i$
    \begin{align}
        L^{(i+1)}=\left\lceil {\frac{\log v_{L^{(i)}}}{\log u}} \right\rceil \geq L^*.
    \end{align}
    If $L^{(i)}>L^*$, then, \eqref{eq::conv_up} implies
    \begin{align}
        L^{(i+1)}=\left\lceil {\frac{\log v_{L^{(i)}}}{\log u}} \right\rceil<L^{(i)}-1. \label{eq::algo_iteration_cov}
    \end{align}
    and by 
    \begin{align}
        L^*\geq \frac{\log v_{L^*}}{\log u} >L^*-1,
    \end{align}
    the result $L^{(i)}$ converges in at most $L^{(2)}-L^*$ loops to $L^*$. It remains to show $L^{(2)}<\infty$ when $L^{*}$ is finite. 
    By \eqref{eq::conv_up}, we have 
    \begin{align}
        L^{(2)}=\min\left\{\left\lceil\frac{\log v_{2}}{\log u}\right\rceil,\left\lceil\frac{\log v_{\infty}}{\log u}\right\rceil \right\}\leq \left\lceil\frac{\log v_{\infty}}{\log u}\right\rceil<\infty,
    \end{align}
    which completes the proof.
    Note that, if there are multiple $L$ with optimal value, then the algorithm outputs the largest of those $L$.
\end{Proof}

We remark here that for the results of Section~\ref{sec::numerical}, we use $10^{-4}$ precision for $\alpha$ and $\gamma$ values and Algorithm~\ref{alg::cap} runs the while loop at most $4$ times.

\section{Proof of Theorem~\ref{thm::quasiconcavity_stealth} and Algorithm~\ref{alg::cap_stealth}}\label{sec::app-s-steal}
Let $w^{(S)}_\sigma(x,y)$ denote the expected value of $w_\sigma$ associated with the cycle $i$ given $A=x$ and $H=y$ and the attacker employs $S$-stealth mining strategy, \textit{i.e.},
\begin{align}
    &w^{(S)}_\sigma(x,y)\nonumber\\&=w_\sigma(\mathbb{E}^{(S)}[x_i|A=x,H=y],\mathbb{E}^{(S)}[y_i|A=x,H=y])\\
    &=(1-\sigma)\mathbb{E}^{(S)}[x_i|A=x,H=y]-\sigma\mathbb{E}^{(S)}[y_i|A=x,H=y].
\end{align}

\begin{lemma}\label{lemma::bounds_strategies_stealth}
If $\sigma_{S+1}<\sigma_S$, then
\begin{align}
    &w^{(S+1)}_{\sigma_{S}}(S-1,S)+\frac{2(2S-1)}{S+1}\frac{\alpha}{1-\alpha} w^{(S+1)}_{\sigma_{S}}(S,S-1) \nonumber\\
    &< w^{(S+1)}_{\sigma_{S+1}}(S-1,S)+\frac{2(2S-1)}{S+1}\frac{\alpha}{1-\alpha} w^{(S+1)}_{\sigma_{S+1}}(S,S-1) \nonumber\\
    &< w^{(S)}_{\sigma_{S}}(S-1,S)+\frac{2(2S-1)}{S+1}\frac{\alpha}{1-\alpha} w^{(S)}_{\sigma_{S}}(S,S-1) \nonumber\\
    &< w^{(S)}_{\sigma_{S+1}}(S-1,S)+\frac{2(2S-1)}{S+1}\frac{\alpha}{1-\alpha} w^{(S)}_{\sigma_{S+1}}(S,S-1). \label{eq:lemma_strategies_relations_stealth}
\end{align}
Similarly, if $\sigma_{S+1}>\sigma_S$, we can switch the direction of the inequalities above.
\end{lemma}

\begin{Proof}
    As usual, $q_{S}$ denotes the probability that a cycle passes through $A=H+1=S$. Further, let $p_{S}$ denote the probability that a cycle passes through $H=A+1=S$. Then,
\begin{align}
    0=&\lim_{T\to\infty}\frac{1}{T}\sum_{i=1}^{T}w^{(S)}_{\sigma_{S}}(x_i,y_i)\\
    =&q_{S}w^{(S)}_{\sigma_{S}}(S,S-1)+p_{S}w^{(S)}_{\sigma_{S}}(S-1,S)\nonumber\\&+(1-q_{S}-p_{S})w^{(S)}_{\sigma_{S}}(\overline{S-1}),\label{eq::a_strategy_stealth}
\end{align}
where $w^{(S)}_{\sigma_{S}}(\overline{S-1})$ is the expected value of $w_\rho$ for $S$-stealth mining strategy given $\overline{S-1}$, \textit{i.e.}, the cycle does not pass through $A=H+1=S$ or $H=A+1=S$. Next, for any $\sigma\in [0,1]$
\begin{align}
    w^{(S)}_{\sigma}(\overline{S-1})=w^{(S+1)}_{\sigma}(\overline{S-1}),
\end{align}
which is due to the fact that both strategies have the same rewards for all possible arrival paths that do not pass through the event $A=H+1=S$ or $H=A+1=S$ during a cycle. Here, unlike $L$-stubborn mining, we need to exclude $H=A+1=S$ as well. To see this, notice that $(S+1)$-stealth mining does not match the honest block at height $S-1$, whereas $S$-stealth mining does, hence their rewards differ and we cannot apply the separation of dimensions argument explained in \figref{fig::coordinate}.

Further, if $\sigma_{S+1}<\sigma_S$, then
\begin{align}
w^{(S)}_{\sigma_{S}}(\overline{S-1})<w^{(S+1)}_{\sigma_{S+1}}(\overline{S-1}).
\label{eq::lvslplus1_stealth}
\end{align}
Thus, rewriting \eqref{eq::a_strategy_stealth} for $(S+1)$-stealth mining strategy
\begin{align}
    0=&q_{S}w^{(S+1)}_{\sigma_{S+1}}(S,S-1)+p_{S}w^{(S+1)}_{\sigma_{S+1}}(S-1,S)\nonumber\\&+(1-q_{S}-p_{S})w^{(S+1)}_{\sigma_{S+1}}(\overline{S-1})\\
    >& q_{S}w^{(S+1)}_{\sigma_{S+1}}(S,S-1)+p_{S}w^{(S+1)}_{\sigma_{S+1}}(S-1,S)\nonumber\\&+(1-q_{S}-p_{S})w^{(S)}_{\sigma_{S}}(\overline{S-1})\\
    =&q_{S}\left(w^{(S+1)}_{\sigma_{S+1}}(S,S-1)-w^{(S)}_{\sigma_{S}}(S,S-1)\right) \nonumber\\&+ p_{S}\left(w^{(S+1)}_{\sigma_{S+1}}(S-1,S)-w^{(S)}_{\sigma_{S}}(S-1,S)\right). \label{eq::a_strategy_proved_stealth}
\end{align}
Notice, a cycle passes through $H=A+1=S$ w.p.
\begin{align}
    p_{S}&=P[S-1,S-1]\alpha^{S-1}\beta^{S}\\
    &=C_{S-1}\alpha^{S-1}\beta^{S},
\end{align}
and through $A=H+1=S$ w.p. 
\begin{align}
    q_{S}&=P[S,S-1]\alpha^{S}\beta^{S-1}\\
    &=C_{S}\alpha^{S}\beta^{S-1}\\
    &=\frac{2(2S-1)}{S+1}\cdot\frac{\alpha}{1-\alpha}\cdot p_{S},
\end{align}
which concludes the proof.
\end{Proof}

\begin{lemma}\label{lemma::eq_strategies_stealth_1}
Given $A=H+1=S$, the expected value of $w_\sigma$ associated with $(S+1)$-stealth mining strategy is related to $S$-stealth mining strategy in the following way,
    \begin{align}
    &w^{(S+1)}_{\sigma}(S,S-1)\nonumber\\&=w^{(S)}_{\sigma}(S,S-1)-\beta^2\left((S+1)-\gamma S-\frac{1-\sigma}{1-2\alpha} \right).\label{eq::narrow-decision_stealth_1}
\end{align}
\end{lemma}

\begin{Proof}
    Clearly, if $S$-stealth mining strategy is followed, we get $ w^{(S)}_{\sigma}(S,S-1)=w_{\sigma}(S,0)$. On the other hand, if $(S+1)$-stealth mining strategy is followed, the decision is delayed and $w^{(S+1)}_{\sigma}(S,S-1)$ depends on the arrival of the next blocks in the following way:
    \begin{enumerate}
    \item The next block is adversarial w.p. $\alpha$. In this case, the adversarial advantage $A-H=2$ implies that the decision will be made when the advantage reduces to $1$, hence w.p. $\alpha$ we get $w_{\sigma}(S+1+\frac{\alpha}{1-2\alpha},0)$.
    \item The next block is honest followed by an adversarial block w.p. $\beta\alpha$.  As a result, we get $w_{\sigma}(S+1,0)$ w.p. $\beta\alpha$.
    \item The next two blocks are honest w.p. $\beta^2$. In this situation, $w_{\sigma}$ depends on whether the last honest block switches prefix or not:
    \begin{enumerate}
        \item The prefix switches to adversarial blocks w.p. $\gamma$ and we have $w_{\sigma}(S,1)$.
        \item The prefix does not contain any adversarial block w.p. $(1-\gamma)$ and we have $w_{\sigma}(0,S+1)$.
    \end{enumerate}
\end{enumerate}
Combining the cases above in expectation, for $(S+1)$-stealth mining strategy, we obtain
\begin{align}
    &w^{(S+1)}_{\sigma}(S,S-1)=\alpha w_{\sigma}(S+1+\frac{\alpha}{1-2\alpha},0)\nonumber\\&+\beta\alpha w_{\sigma}(S+1,0)+\beta^2\gamma w_{\sigma}(S,1)\nonumber\\&+\beta^2(1-\gamma)w_{\sigma}(0,S+1),
\end{align}
which results in \eqref{eq::narrow-decision_stealth_1}.
\end{Proof}

\begin{lemma}\label{lemma::eq_strategies_stealth_2}
Given $A-1=H=S$, the expected value of $w_\sigma$ associated with $(S+1)$-stealth mining strategy is related to $S$-stealth mining strategy in the following way,
    \begin{align}
    w^{(S+1)}_{\sigma}(S-1,S)=w^{(S)}_{\sigma}(S-1,S)-\gamma(S-1).\label{eq::narrow-decision_stealth_2}
\end{align}
\end{lemma}
\begin{Proof}
    Clearly, if $(S+1)$-stealth mining strategy is followed, we get $ w^{(S+1)}_{\sigma}(S-1,S)=w_\sigma(0,S)=-\sigma S$. On the other hand, if $S$-stealth mining strategy is followed, the reward depends on the prefix of the last honest block:
\begin{enumerate}
    \item The prefix contains no adversarial blocks w.p. $1-\gamma$. In this case, we get $w_\sigma(0,S)$.
    \item The prefix contains $S-1$ adversarial blocks w.p. $\gamma$. In this case, we get $w_\sigma(S-1,1)$.
\end{enumerate}
Combining the cases above results in \eqref{eq::narrow-decision_stealth_2}.
\end{Proof}

The intermediary results above are now sufficient to prove Theorem~\ref{thm::quasiconcavity_stealth}.

If $\sigma_{S}\geq\sigma_{S+1}$, picking $\sigma=\sigma_{S+1}$ in \eqref{eq::narrow-decision_stealth_1} and \eqref{eq::narrow-decision_stealth_2} implies
\begin{align}
    \gamma\frac{S^2-1}{2(2S-1)}+\alpha\beta(S+1-\gamma S)\geq \alpha\beta\frac{1-\sigma_{S+1}}{1-2\alpha},\label{eq::lemmacombine_stealth_1}
\end{align}
due to Lemma \ref{lemma::bounds_strategies_stealth}.
 
If $\sigma_{S+1}<\sigma_{S+2}$, rewriting Lemma~\ref{lemma::bounds_strategies_stealth} for the relation between $(S+1)$ and $(S+2)$-stealth mining and picking $\sigma=\sigma_{S+1}$ in \eqref{eq::narrow-decision_stealth_1} and \eqref{eq::narrow-decision_stealth_2}, we get
\begin{align}
    \alpha\beta\frac{1-\sigma_{S+1}}{1-2\alpha}>\gamma\frac{S(S+2)}{2(2S+1)}+\alpha\beta(S+2-\gamma(S+1)). \label{eq::lemmacombine_stealth}
\end{align}
Combining the two implies,
\begin{align}
    0>\gamma\frac{2S^2+1}{4S^2-1}+2\alpha\beta(1-\gamma),
\end{align}
which creates a contradiction since the right hand side is nonnegative for $S\geq 1$, $\beta \geq 0.5\geq \alpha\geq 0$ and $1\geq \gamma\geq 0$. This concludes the proof of Theorem \ref{thm::quasiconcavity_stealth}.

It is clear that $f(\sigma)<\infty$ for $\sigma \in[0,1]$. Further, one can also show that 
\begin{align}
    S^{*}\geq f(\sigma_{S^{*}})>S^{*}-1,
\end{align}
by Theorem~\ref{thm::quasiconcavity_stealth}. This in turn, by modifying the arguments of Proposition~\ref{prop::self} with the results proven in this section, implies that Algorithm~\ref{alg::cap_stealth} returns $S^{*}$. 

\bibliographystyle{ieeetr}
\bibliography{blockchain}

\begin{thebibliography}{10}

\bibitem{btc-whitepaper}
S.~Nakamoto, ``Bitcoin: A peer-to-peer electronic cash system.'' https://bitcoin.org/bitcoin.pdf, March 2008.

\bibitem{blockchain_security_survey}
X.~Li, P.~Jiang, T.~Chen, X.~Luo, and Q.~Wen, ``A survey on the security of blockchain systems,'' {\em Future Generation Computer Systems}, vol.~107, pp.~841--853, 2020.

\bibitem{rosenfeld2014analysis}
M.~Rosenfeld, ``Analysis of hashrate-based double spending,'' 2014.

\bibitem{selfish-mining}
I.~Eyal and E.~G. Sirer, ``Majority is not enough: Bitcoin mining is vulnerable,'' {\em Communications of the ACM}, vol.~61, p.~95–102, July 2018.

\bibitem{optimal-selfish}
A.~Sapirshtein, Y.~Sompolinsky, and A.~Zohar, ``Optimal selfish mining strategies in bitcoin,'' in {\em Springer FC}, 2017.

\bibitem{stubborn-mining}
K.~Nayak, S.~Kumar, A.~Miller, and E.~Shi, ``Stubborn mining: Generalizing selfish mining and combining with an eclipse attack,'' in {\em IEEE EuroS{\&}P}, March 2016.

\bibitem{on_the_sec_pow_gervais}
A.~Gervais, G.~O. Karame, K.~W\"{u}st, V.~Glykantzis, H.~Ritzdorf, and S.~Capkun, ``On the security and performance of proof of work blockchains,'' in {\em ACM CCS}, October 2016.

\bibitem{sompolinsky2016bitcoin}
Y.~Sompolinsky and A.~Zohar, ``Bitcoin's security model revisited,'' 2016.

\bibitem{prob-selfish-mdp-method}
R.~Zur, I.~Eyal, and A.~Tamar, ``Efficient mdp analysis for selfish-mining in blockchains,'' in {\em ACM AFT}, October 2020.

\bibitem{preneel_common_metrics}
R.~Zhang and B.~Preneel, ``Lay down the common metrics: Evaluating proof-of-work consensus protocols' security,'' in {\em IEEE SP}, 2019.

\bibitem{MDP_proof-systems-selfish}
K.~Chatterjee, A.~Ebrahimzadeh, M.~Karrabi, K.~Pietrzak, M.~Yeo, and D.~Zikelic, ``Fully automated selfish mining analysis in efficient proof systems blockchains,'' in {\em ACM PODC}, June 2024.

\bibitem{keller2024genericselfishminingmdp}
P.~Keller, ``Generic selfish mining mdp for dag protocols,'' 2024.

\bibitem{yang2021-deep-dive-analysis-selfish-stubborn}
R.~Yang, X.~Chang, J.~Mišić, and V.~B. Mišić, ``Deep-dive analysis of selfish and stubborn mining in bitcoin and ethereum,'' 2021.

\bibitem{optimal_stubborn_state_transition}
Y.~Zhang, M.~Zhao, T.~Li, Y.~Wang, and T.~Liang, ``Achieving optimal rewards in cryptocurrency stubborn mining with state transition analysis,'' {\em Inf. Sci.}, vol.~625, no.~C, p.~299–313, 2023.

\bibitem{optimal_stubborn_mdp}
Y.~Zhang, M.~Liu, J.~Guo, Z.~Wang, Y.~Wang, T.~Liang, and S.~Singh, ``Optimal revenue analysis of the stubborn mining based on markov decision process,'' in {\em ML4CS}, Springer-Verlag, December 2022.

\bibitem{profitable_double_sp}
J.~Jang and H.-N. Lee, ``Profitable double-spending attacks,'' {\em Applied Sciences}, vol.~10, no.~23, 2020.

\bibitem{Grunspan_profitable_double_sp}
C.~Grunspan and R.~Pérez-Marco, ``On profitability of nakamoto double spend,'' {\em Probability in the Engineering and Informational Sciences}, vol.~36, no.~3, p.~732–746, 2022.

\bibitem{courtois2014subversiveminerstrategiesblock}
N.~T. Courtois and L.~Bahack, ``On subversive miner strategies and block withholding attack in bitcoin digital currency,'' 2014.

\bibitem{miners_dilemma}
I.~Eyal, ``The miner's dilemma,'' in {\em IEEE S\&P}, May 2015.

\bibitem{fork_after_witholding_attack}
Y.~Kwon, D.~Kim, Y.~Son, E.~Vasserman, and Y.~Kim, ``Be selfish and avoid dilemmas: Fork after withholding (faw) attacks on bitcoin,'' in {\em ACM SIGSAC CCS}, October 2017.

\bibitem{power_adjusting}
S.~Gao, Z.~Li, Z.~Peng, and B.~Xiao, ``Power adjusting and bribery racing: Novel mining attacks in the bitcoin system,'' in {\em ACM SIGSAC CCS}, November 2019.

\bibitem{selfholding}
X.~Dong, F.~Wu, A.~Faree, D.~Guo, Y.~Shen, and J.~Ma, ``Selfholding: A combined attack model using selfish mining with block withholding attack,'' {\em Computers \& Security}, vol.~87, p.~101584, 2019.

\bibitem{intermittent_mining}
K.~A. Negy, P.~R. Rizun, and E.~G. Sirer, ``Selfish mining re-examined,'' in {\em Springer FC}, 2020.

\bibitem{Grunspan_witholding_resilience}
C.~Grunspan and R.~P\'{e}rez-Marco, ``Block withholding resilience,'' {\em Digital Finance}, Feb 2025.

\bibitem{profit_lag}
C.~Grunspan and R.~P{\'e}rez-Marco, ``Profit lag and alternate network mining,'' in {\em Springer MARBLE}, pp.~115--132, 2023.

\bibitem{time_average_selfish_mining}
R.~Sarenche, R.~Zhang, S.~Nikova, and B.~Preneel, ``Selfish mining time-averaged analysis in bitcoin: Is orphan reporting an effective countermeasure?,'' {\em IEEE Transactions on Information Forensics and Security}, vol.~20, pp.~449--464, 2025.

\bibitem{freshness_preferred}
E.~Heilman, ``One weird trick to stop selfish miners: Fresh bitcoins, a solution for the honest miner,'' in {\em Financial Cryptography and Data Security}, Springer Berlin Heidelberg, March 2014.

\bibitem{Preventing_Selfish_Creation_Time}
J.~Lee and Y.~.Kim, ``Preventing bitcoin selfish mining using transaction creation time,'' in {\em IEEE ICSSA}, July 2018.

\bibitem{Publish_or_Perish}
R.~Zhang and B.~Preneel, ``Publish or perish: A backward-compatible defense against selfish mining in bitcoin,'' in {\em CT-RSA}, Springer International Publishing, February 2017.

\bibitem{Gobel_selfish_mine_prop_delay}
J.~Göbel, H.~Keeler, A.~Krzesinski, and P.~Taylor, ``Bitcoin blockchain dynamics: The selfish-mine strategy in the presence of propagation delay,'' {\em Performance Evaluation}, vol.~104, pp.~23--41, 2016.

\bibitem{Impact_of_Network_Connectivity_on_Consensus}
Y.~Xiao, N.~Zhang, W.~Lou, and Y.~T. Hou, ``Modeling the impact of network connectivity on consensus security of proof-of-work blockchain,'' in {\em IEEE INFOCOM}, July 2020.

\bibitem{survey_double_sp_selfish}
K.~Nicolas, Y.~Wang, G.~C. Giakos, B.~Wei, and H.~Shen, ``Blockchain system defensive overview for double-spend and selfish mining attacks: A systematic approach,'' {\em IEEE Access}, vol.~9, pp.~3838--3857, 2021.

\bibitem{ballot-problem}
J.~Bertrand, ``Solution d'un problème,'' {\em Comptes Rendus de l'Académie des Sciences de Paris}, vol.~105, p.~369, 1887.

\bibitem{baileys_number}
D.~F. Bailey, ``Counting arrangements of 1's and -1's,'' {\em Mathematics Magazine}, vol.~69, no.~2, pp.~128--131, 1996.

\bibitem{catalan-numbers-book}
T.~Koshy, {\em {Catalan Numbers with Applications}}.
\newblock Oxford University Press, 11 2008.

\bibitem{blockchains_biased_ballot_problem}
L.~Chen, L.~Xu, Z.~Gao, N.~Shah, Y.~Lu, and W.~Shi, ``{Smart Contract Execution - the (+-)-Biased Ballot Problem},'' in {\em ISAAC}, July 2017.

\bibitem{catalan-stubborn-grunspan}
C.~Grunspan and R.~P\'{e}rez-Marco, ``{Selfish Mining and Dyck Words in Bitcoin and Ethereum Networks},'' in {\em Tokenomics}, May 2019.

\bibitem{grunspan2020mathematics}
C.~Grunspan and R.~Pérez-Marco, ``The mathematics of bitcoin,'' {\em European Mathematical Society - Newsletter}, vol.~115, p.~31–37, 2020.

\bibitem{Goffard_Fraud_risk}
P.-O. Goffard, ``Fraud risk assessment within blockchain transactions,'' {\em Advances in Applied Probability}, vol.~51, no.~2, p.~443–467, 2019.

\bibitem{grunspan2019-profitability-selfish-mining}
C.~Grunspan and R.~P\'{e}rez-Marco, ``On profitability of selfish mining,'' 2019.

\bibitem{werlman}
R.~Bar-Zur, A.~Abu-Hanna, I.~Eyal, and A.~Tamar, ``Werlman: to tackle whale (transactions), go deep (rl),'' in {\em ACM SYSTOR}, June 2022.

\bibitem{Deep_Bribe}
R.~Bar-Zur, D.~Dori, S.~Vardi, I.~Eyal, and A.~Tamar, ``{Deep Bribe: Predicting the Rise of Bribery in Blockchain Mining with Deep RL},'' in {\em IEEE SPW}, May 2023.

\bibitem{squirrl}
C.~Hou, M.~Zhou, Y.~Ji, P.~Daian, F.~Tramer, G.~Fanti, and A.~Juels, ``Squirrl: Automating attack analysis on blockchain incentive mechanisms with deep reinforcement learning,'' in {\em NDSS Symposium}, February 2021.

\bibitem{nakamoto-always-wins}
A.~Dembo, S.~Kannan, E.~Tas, D.~Tse, P.~Viswanath, X.~Wang, and O.~Zeitouni, ``Everything is a race and {N}akamoto always wins,'' in {\em ACM CCS}, November 2020.

\bibitem{cao2023tradeoff}
S.-J. Cao and D.~Guo, ``Security, latency, and throughput of proof-of-work nakamoto consensus,'' {\em IEEE Transactions on Information Theory}, vol.~71, no.~6, pp.~4708--4731, 2025.

\bibitem{our-sec-lat-extended}
M.~Doger and S.~Ulukus, ``Refined bitcoin security-latency under network delay,'' {\em IEEE Transactions on Information Theory}, vol.~71, no.~4, pp.~3038--3047, 2025.

\bibitem{our-queue-sec-ext-version}
M.~Doger and S.~Ulukus, ``Transaction capacity, security and latency in blockchains,'' 2025.

\bibitem{Sompolinsky2015SecureHT}
Y.~Sompolinsky and A.~Zohar, ``Secure high-rate transaction processing in bitcoin,'' in {\em Financial Cryptography}, 2015.

\bibitem{PHANTOM_GHOSTDAG}
Y.~Sompolinsky, S.~Wyborski, and A.~Zohar, ``Phantom ghostdag: a scalable generalization of nakamoto consensus: September 2, 2021,'' in {\em ACM AFT}, September 2021.

\bibitem{SPECTRE}
Y.~Sompolinsky, Y.~Lewenberg, and A.~Zohar, ``{SPECTRE}: A fast and scalable cryptocurrency protocol.'' Cryptology {ePrint} Archive, Paper 2016/1159, 2016.

\bibitem{Impact_of_Temporary_Fork}
C.~Chen, X.~Chen, J.~Yu, W.~Wu, and D.~Wu, ``Impact of temporary fork on the evolution of mining pools in blockchain networks: An evolutionary game analysis,'' {\em IEEE Transactions on Network Science and Engineering}, vol.~8, no.~1, pp.~400--418, 2021.

\bibitem{sakurai2024modelbasedanalysisminingfairness}
S.~Akira and S.~Kazuyuki, ``Model-based calculation method of mining fairness in blockchain,'' {\em IEEE Open Journal of the Computer Society}, vol.~7, pp.~129--141, 2026.

\bibitem{mitsuhamizu}
Mitsuhamizu, ``A python implementation for solving the mdp in optimal selfish mining.''

\bibitem{andre-ballot-solution}
D.~Andre, ``Solution directe du probleme resolu par m. bertrand,'' {\em Comptes Rendus de l'Académie des Sciences de Paris}, vol.~105, pp.~436--437, 1887.

\end{thebibliography}

\end{document}